\documentclass[twocolumn]{aa}
\usepackage{graphicx}

\newcommand{\lstar}{L^{\displaystyle\ast}_X}
\newcommand{\phistar}{\phi^{\displaystyle\ast}}
\newcommand{\be}{\begin{equation}}
\newcommand{\ee}{\end{equation}}
\def\ltsima{$\; \buildrel < \over \sim \;$}
\def\lsim{\lower.5ex\hbox{\ltsima}}
\def\gtsima{$\; \buildrel > \over \sim \;$}\def\gsim{\lower.5ex\hbox{\gtsima}}
\def\deg {^\circ}

\begin{document}

\title{The Brera Multi-scale Wavelet HRI Cluster Survey: 
       I Selection of the sample and number counts%
\thanks{Partially based on observations taken at ESO and TNG telescopes.}}

\author{A. Moretti\inst{1}, L. Guzzo\inst{1}, S. Campana\inst{1},
D. Lazzati\inst{2}, M.R. Panzera\inst{1}, 
G. Tagliaferri\inst{1}, 
S. Arena\inst{1}, F. Braglia\inst{1}, 
I. Dell'Antonio\inst{3}, M.~Longhetti \inst{1}
}
\offprints{moretti@merate.mi.astro.it}

\institute{INAF Osservatorio Astronomico di Brera, Via E. Bianchi
46, Merate (LC), 23807, Italy
\and
Institute of Astronomy, University of Cambridge, 
Madingley Road, Cambridge CB3 0HA, UK.
\and 
Brown University, Providence, RI 02912, USA}

\date{Received ; accepted }

\titlerunning{The BMW--HRI Cluster Survey}
\authorrunning{Moretti et al.\ }

\abstract{
We describe the construction of the Brera Multi-scale Wavelet (BMW) HRI Cluster Survey, a deep
sample of serendipitous X--ray selected clusters of galaxies based on
the ROSAT HRI archive.  This is the first cluster catalog exploiting
the high angular resolution of this instrument.  Cluster candidates
are selected on the basis of their X--ray extension only, a parameter
which is well measured by the BMW wavelet detection algorithm.  The
survey includes 154 candidates over a total solid angle of $\sim$ 160
deg$^2$ at 10$^{-12}$ erg s$^{-1}$ cm$^{-2}$ and $\sim$ 80 deg$^2$ at
1.8$\times$10$^{-13}$ erg s$^{-1}$ cm$^{-2}$.  At the same time, a
fairly good sky coverage in the faintest flux bins ($3-5
\times$10$^{-14}$ erg s$^{-1}$ cm$^{-2}$) gives this survey the capability of
detecting a few clusters with $z\sim 1-1.2$, depending on evolution.  We
present the results of extensive Monte Carlo simulations, providing a
complete statistical characterization of the survey selection function
and contamination level.  
We also present a new estimate of the surface density of clusters of
galaxies down to a flux of $3\times$10$^{-14}$ erg s$^{-1}$
cm$^{-2}$, which is consistent with previous measurements from PSPC-based
samples. Several clusters with redshifts up to $z=0.92$ have
already been confirmed, either by cross-correlation with existing PSPC
surveys or from early results of an ongoing follow-up campaign.
Overall, these results indicate that the excellent HRI PSF (5\arcsec
~FWHM on axis) more than compensates for the negative effect of the
higher instrumental background on the detection of
high-redshift clusters.  In addition, it allows us to detect
compact clusters that could be lost at lower resolution, thus
potentially providing an important new insight into cluster evolution.
}
\maketitle

\section{Introduction}
\begin{figure*}
  \centering
\includegraphics[width=16cm]{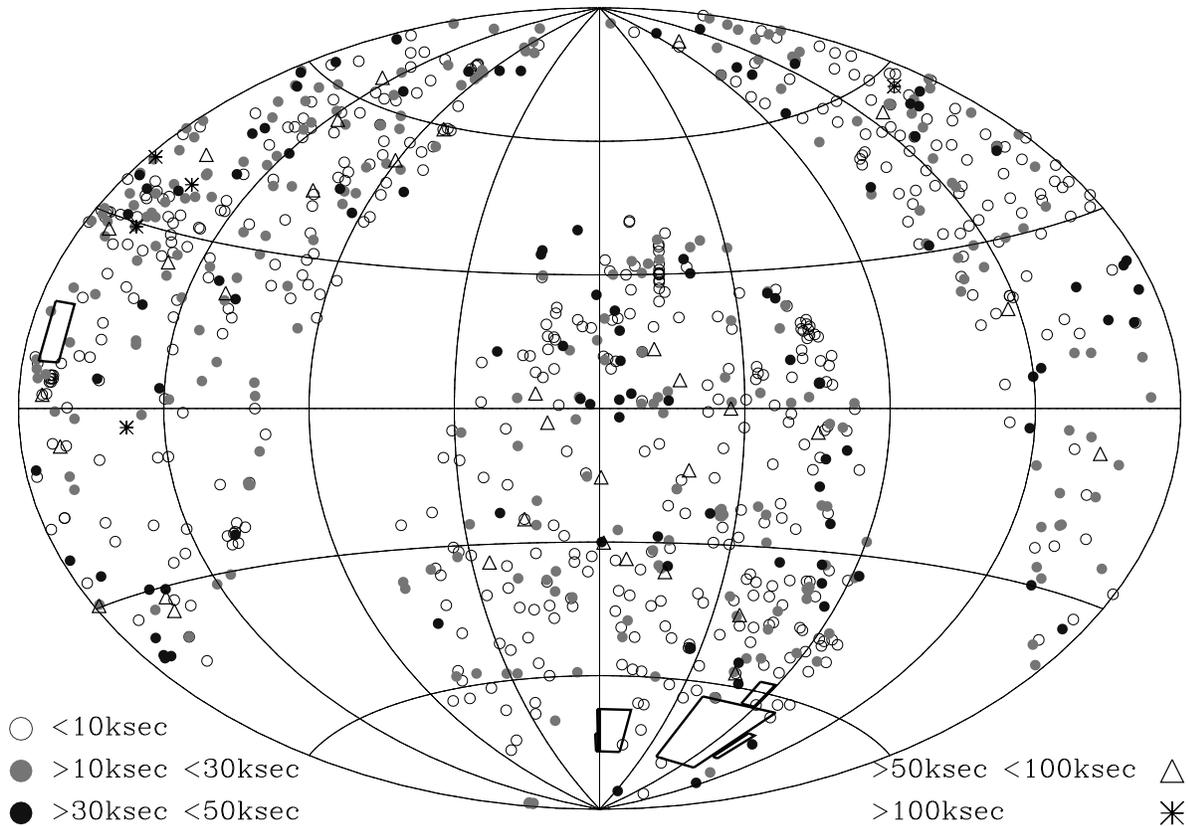}
\caption{The spatial distribution (equatorial coordinates, Aitoff projection) of the 914 HRI 
pointings used for the BMW--HRI cluster survey.  The origin of right
ascension is in the center of the plot and the grid is in steps of $3^h$
in RA and $30^\circ$ in DEC.  The empty area corresponds to the
{$|b_{II}| < 20^\circ$} zone of avoidance, while the boxes mark the Magellanic clouds and
Virgo cluster areas which were also excluded from the survey.  Different
symbols indicate different ranges of exposure time, as explained.}
\label{map} 
\end{figure*}
Clusters of galaxies represent the largest collapsed objects in the
hierarchy of cosmic structures, resulting from the growth of fluctuations
lying on the high--density tail of the matter density field
(e.g. Peacock 1999).  As such, their number density and evolution are
strongly dependent on the normalization of the power spectrum and the
value of the density parameter $\Omega_M$ (e.g. Borgani \& Guzzo
2001; Rosati et al. 2002).  In addition, clusters are usually considered as ``simple''
systems, where the physics involved in turning ``mass into light'' is
possibly easier to understand, compared to the various complex processes
connected to star formation and evolution in galaxies.  In particular
in the X--ray band, where clusters can be defined and recognized as
single objects (not just as a mere collection of galaxies), observable quantities
like X--ray luminosity $L_X$ and temperature $T_X$ show fairly tight
relations with the cluster mass (e.g. Evrard et al. 1996; Allen et
al. 2001; Reiprich \& B\"ohringer 2002; Ettori et al. 2004).  A
full comprehension of these scaling relations requires more
ingredients than a simple heating during the
growth of fluctuations (Kaiser 1986; Helsdon \& Ponman 2000; Finoguenov et
al. 2001; Borgani et al. 2004).  However, their existence and relative tightness give clusters a
specific role as probes of the cosmological model, providing us with a way to test
fairly directly the mass function (e.g. B\"ohringer et al. 2002;
Pierpaoli et al. 2003) and the mass power spectrum (Schuecker et al. 2003), via respectively
the observed cluster X--ray luminosity function (XLF) and clustering.  In addition, and
equally important, clusters at different redshifts provide homogeneous samples of essentially
coeval galaxies in a high-density environment and give the possibility of studying the
evolution of stellar populations (e.g. Blakeslee et al. 2003; Lidman
et al. 2004).

X--ray based cluster surveys, in addition to a fairly direct connection of the 
observed quantities to model (mass-specific) predictions,  have one
further, fundamental advantage over optically-selected catalogues: their
selection function is well-defined (being essentially that of a flux-limited
sample) and fairly easy to reconstruct.   This is a crucial feature when the goal is to use
these samples for cosmological measurements that necessarily involve a
precise knowledge of the sampled volume, as is the case when
computing first or second moments of the density field. 
X--ray surveys in the ``local'' Universe ($ \left< z\right> \sim 0.1
$), stemming from the ROSAT All-Sky Survey (RASS, Voges et
al. 1999) have been able to pinpoint quantities like the cluster
number density to high accuracy.  The REFLEX survey, in particular,
has yielded the currently most accurate measurement of the XLF (B\"ohringer
et al. 2002; 2004), in substantial agreement with other local estimates
(Ebeling et al. 1997; De Grandi et al. 2001).  These result provide
a robust $z\sim 0$ reference frame to which surveys of distant clusters can
be safely compared in search of evolution.  

Recent X--ray searches for serendipitous high-redshift
clusters have been based mostly on the deeper pointed images collected
with the ROSAT PSPC instrument (RDCS: Rosati et al. 1995; SHARC:
Collins et al. 1997; 160 Square Degrees: Vikhlinin et al. 1998a;
WARPS: Perlman et al. 2002) or on the high-exposure North Ecliptic
Pole area of the RASS (NEP: Gioia et al. 2003); a deeper search for
massive clusters in the overall RASS is also being carried out (MACS, Ebeling et
al. 2001a). Results from these surveys consistently show a lack of evolution of the XLF\footnote{All
through this paper, we shall adopt a ``concordance'' cosmological
model, with $H_o=70$ km s$^{-1}$ Mpc$^{-1}$, $\Omega_M=0.3$,
$\Omega_{\Lambda} = 0.7$, and --- unless specified --- quote all X--ray
fluxes and luminosities in the [0.5-2] keV band.} for $L <
\lstar\simeq 3\cdot 10^{44}$ erg s$^{-1}$ out to $z\sim 0.8$, pointing to
low values of $\Omega_M$ under reasonable assumptions on the evolution
of the $L_X-T_X$ relation (see Rosati et al. 2002 for a review).  At the same time,
however, they confirm the early findings from the Einstein Medium
Sensitivity Survey (EMSS, Gioia et al. 1990; Henry et al. 1992) of a
mild evolution of the bright end (Vikhlinin et al. 1998a; Nichol et
al. 1999; Borgani et al. 2001; Gioia et al. 2001; Mullis et al. 2004). In
other words, there is an indication that above $z \sim 0.6$ one finds
less very massive clusters, likely indicating that beyond this epoch
they were still to be assembled from the merging of smaller mass
units.
These conclusions, however, are still based on a rather small number
of high--redshift clusters.  Currently, we know $\sim
15$ X--ray confirmed clusters above $z=0.8$, with only 5 so far detected above
$z=1$, all of which are at $z<1.3$.   In addition, virtually all
current statistical samples
of distant clusters have been selected from X--ray images collected
with the same, low angular resolution ROSAT-PSPC instrument.  Clearly,
the need for more samples of high-redshift clusters, possibly selected
from independent X--ray imaging material, is compelling.

As a first contribution in this direction, during the last four years
we have constructed a new sample of distant cluster candidates based
on the still unexplored ROSAT-HRI archive, the BMW--HRI Cluster Survey.
While serendipitous searches are already focusing on the fresh data
being accumulated in the archives of the new powerful X--ray satellites
XMM-Newton and Chandra (see e.g. the pioneering work by Boschin
2002), our work shows that a remarkable source of high-redshift
clusters --- the HRI archive --- has so far been neglected.  In this paper
we discuss in detail the selection process, the properties of this
catalogue, its completeness function and contamination.  We then
compute the sky coverage, number counts and expected redshift
distribution of BMW--HRI clusters.  The paper is organized as
follows. Sect. 2 describes the general features of the HRI data and
the field selection; Sect. 3 discusses the cluster detection and
characterization; Sect. 4 presents the results of
extensive Monte Carlo simulations, performed in order to understand
the statistical properties of the cluster sample, in particular its
completeness and contamination level; Sect. 5 uses all this
information to derive the survey sky--coverage as a function of source
flux and extension, and to compute the survey expected redshift
distribution and mean surface density; Sect. 6 presents a few
examples of distant clusters already identified in the BMW--HRI survey, while the
last section summarizes the main results obtained in the paper.

\section{Survey Description}
\subsection{The HRI instrument}
The High-Resolution Imager (HRI) was the secondary instrument on board
the ROSAT satellite, with technical features quite different from and
complementary to the companion PSPC, including in particular a much
better spatial resolution.  The core of the HRI is a micro-channel
plate detector with an octagon--like shaped field of view (with $\sim$
19\arcmin~radius) that reveals single X--ray photons, providing
information on their positions and arrival times.  The HRI Point
Spread Function (PSF) as measured on--axis is about 5\arcsec~FWHM,
i.e. a factor of $\sim 4$ better than the PSPC~\footnote{For
completeness, note that this is still a factor of $\sim 5$ worse than
that of the Chandra X--ray Observatory, whose archive represents a
further source of high-resolution data that is currently also under
scrutiny using our algorithms (Romano et al., in preparation)}.  On
the other hand, the HRI is less efficient than the PSPC (a factor of 3
to 8 for a plausible range of incident spectra) and has a higher
background.  This consists of several components: the
internal background due to the residual radioactivity of the detector
(1--2 cts s$^{-1}$), the externally-induced background from charged
particles (1--10 cts s$^{-1}$) and the X--ray background (0.5--2 cts
s$^{-1}$).

\begin{figure} 
\includegraphics[width=8cm]{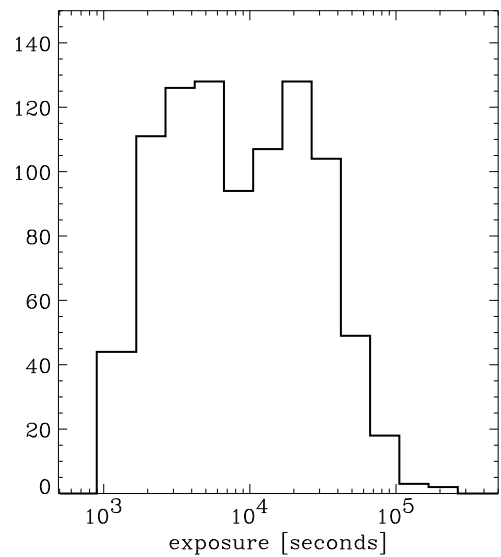} 
\caption{Distribution of exposure times for the 914 pointings used for
the cluster survey. }
\label{skyt} 
\end{figure}

The HRI covers the energy range [0.1--2.4] keV, divided into 16 Pulse
Height Analyzer (PHA) channels, which provide very crude spectral
information (Prestwich et al. 1996). The HRI background is
highest in the first few (1--3) PHA channels and, at variance with the
ROSAT PSPC, it is dominated by the unvignetted particle background. As
shown by sky calibration sources, most of the source photons instead
arrive in a PHA channel between 3 and 8 (David et al. 1998).
A priori, the high background of the HRI can be a serious problem for
cluster searches.  The low surface brightness of these diffuse
objects  is indeed further suppressed as a function of redshift, due to the
cosmological dimming $\propto (1+z)^4$ and rapidly drops below the
background.  Coupled to the HRI lower sensitivity, this explains
why all ROSAT serendipitous cluster searches have so far concentrated
on the PSPC archive (see Sect. 1).  We shall show
here how these early, pessimistic conclusions are fortunately not
fully correct.   
As detailed in Campana et al. (1999, C99 hereafter), in order to minimize the background and increase
the signal-to-noise (S/N) ratio of  
X--ray sources, especially low-surface-brightness ones, the BMW
wavelet analysis has been restricted to PHA 2--9 (see also David et
al. 1998). This range reduces the detector
background by about 40\% with a minimum loss of cosmic X--ray photons
($< 10\%$; David et al. 1998).  This is certainly one key feature in 
improving the detection of clusters of galaxies with these data,
thus allowing studies to reap the benefit of the excellent resolution of this instrument.

\subsection{The surveyed area}
\label{multi-area}
The overall BMW--HRI serendipitous catalogue (Panzera et al. 2003, P03
hereafter; Lazzati et al. 1999, L99 hereafter; C99)
is based on the analysis of 4298 observations, i.e. the
whole HRI archive after excluding calibration observations, fields
pointed at supernova remnants and some other problematic pointings.
The selection of cluster candidates imposes a number of extra
selections, which are discussed here.  We first selected only high
Galactic latitude fields ($ |b| > 20^{\circ}$) with exposure times
larger than 1000 seconds, excluding also the Magellanic Clouds and
Virgo Cluster regions, defined in Table~\ref{excluded-areas}.
Moreover, we excluded all pointings on star clusters, clusters and
groups of galaxies and obviously Messier and NGC nearby galaxies.  In
all these cases the target tends to fill the field-of-view of the
instrument, thus biasing any search for serendipitous sources in the
surrounding area.  For this reason, we further visually inspected
all the remaining fields using panoramic ($40^\prime \times
40^\prime$) DSS2 images with the HRI X--ray contours overlaid, so as to
ascertain whether the central target could in any way influence the
searched region (as e.g. in the case of a spiral galaxy with faint arms
extending out of the central 3-arcmin radius area).  All fields with
such ``problematic'' targets were excluded from the cluster candidate
catalogue.  

   \begin{table}                                                                
      \caption{Regions around the LMC, SMC and Virgo cluster excluded from the survey.}   
      \[                                                                        
         \begin{array}{lll}                                                    
            \hline                                                              
            \noalign{\smallskip}                                                
 {\rm region}& {\rm RA }  & {\rm DEC } \\        
            \noalign{\smallskip}                                                
            \hline                                                              
            \noalign{\smallskip}                                                
{\rm LMC 1}  &  [3^h\, 52^\prime,  6^h \, 52^\prime]  & [-77\deg,   -63\deg]    \\                     
{\rm LMC 2}  &  [5^h\, 24^\prime, 5^h\, 56^\prime]  & [-63\deg, -58\deg]    \\                      
{\rm LMC 3}  & [6^h\, 52^\prime,  7^h\, 12^\prime] & [-74\deg,  -68\deg]  \\                     
{\rm SMC 1}  & [23^h\, 52^\prime,  1^h\,  20^\prime] & [-77 \deg  , -67.5\deg] \\                     
{\rm SMC 2}  & [23^h\,  46 ^\prime,  23^h\, 54^\prime] & [-77\deg  , -73\deg] \\                   
{\rm SMC 3}  & [01^h\,  20^\prime, 02 ^h\, 00^\prime]  & [-72\deg  , -67.5\deg] \\                     
{\rm Virgo}  & [12^h\,  20^\prime,   12^h\, 44^\prime]  & [7.0 \deg  ,  16\deg]  \\                     
            \noalign{\smallskip}                                                
            \hline                                                              
         \end{array}                                                            
      \]                                                                        
\label{excluded-areas}
\end{table}                                                                  

One further problem needing a careful treatment concerns multiple
observations of the same sky areas.  All these multiple pointings have
been analyzed separately by the BMW--HRI general survey: consequently, the
catalogue lists every source detected in each single
observation, including multiple detections of the same source.   To
simplify the treatment of such cases, when multiple observations looked
``exactly'' at the same region of sky, defined as images
with nominal aim points separated by less than 30\arcsec, we discarded
all but the deepest exposure.  However, if the separation was larger
than this, we kept all the (usually overlapping) fields.  The goal here was to maximize 
the survey area, which is the main factor in finding distant luminous
clusters.   These overlaps have been properly treated in the
computation of the actual survey sky coverage, as explained in detail in
Sect.~\ref{skycov}. 

Within each selected field we then considered only sources detected in
the area comprised between 3\arcmin~and 15\arcmin~off--axis angle.
This excludes the central part ($<$ 3\arcmin) of the field of view,
which is normally where the target is pointed, and the very
external part ($>$ 15\arcmin), where the sensitivity drops and the PSF worsens.
We ended up with a grand total of 914 HRI observations
(Fig.\ref{map}) with exposure times ranging from 1 ks to 204 ks 
(Fig.\ref{skyt}), for a total surveyed area of 160.4 square degrees,
$\sim 7 \%$ of which were observed twice or more.

\section {Selection of Cluster Candidates}

\subsection{Source Extension Criterion}
\label{crite}

\begin{figure} 
\includegraphics[width=8cm]{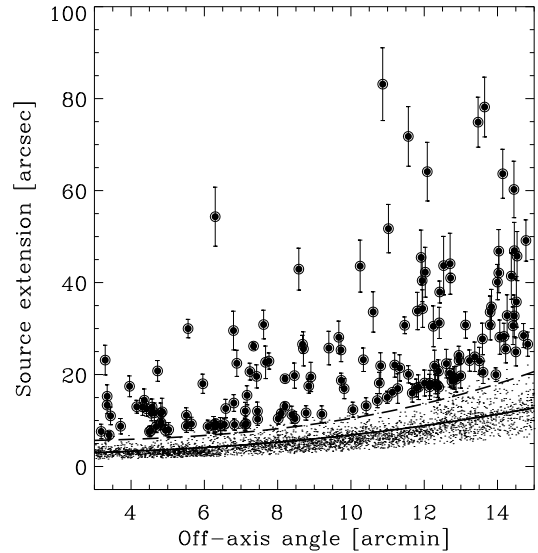}
\caption{Source extensions as measured by the BMW detection algorithm 
(W$_{clu}$, see text),
plotted against the off-axis angle, showing our empirical
definition of extended sources.   The continuous line 
gives the mean extension of a point--like source at different off-axis
angles, and the dashed lines identifies the $3\,\sigma$ locus.  Sources
selected as cluster candidates have to lie, with their error bars,
above the dashed line and are indicated by filled circles.}	
\label{sexten} 
\end{figure}

The BMW detection is based on a wavelet transform (WT) 
algorithm for the automatic detection and characterization of sources
in X--ray images. It is fully described in L99 and has been developed
to analyze ROSAT HRI images, producing the BMW--HRI catalogue (C99,
P03). Candidate sources are identified as local maxima above a given
threshold in wavelet space; the preliminary product of the detection procedure  
is a source list with a rough determination of the position (the center
of the pixel with higher coefficient), source size (the scale of the WT
where the signal to noise is maximized) and total number of photons 
(the value of the maximum WT coefficient). 
Since the wavelet scale steps are discrete,
the final catalogue parameters are a refinement of these values and are
obtained through $\chi^2$ minimization 
with respect to a Gaussian model in WT space (for all the details of
the fit procedure in the WT space see L99).
In particular, the BMW catalog source extension parameter (hereafter W$_{clu}$) 
is defined as the width of the best-fitting Gaussian.

\begin{figure*}  
\includegraphics[width=16cm]{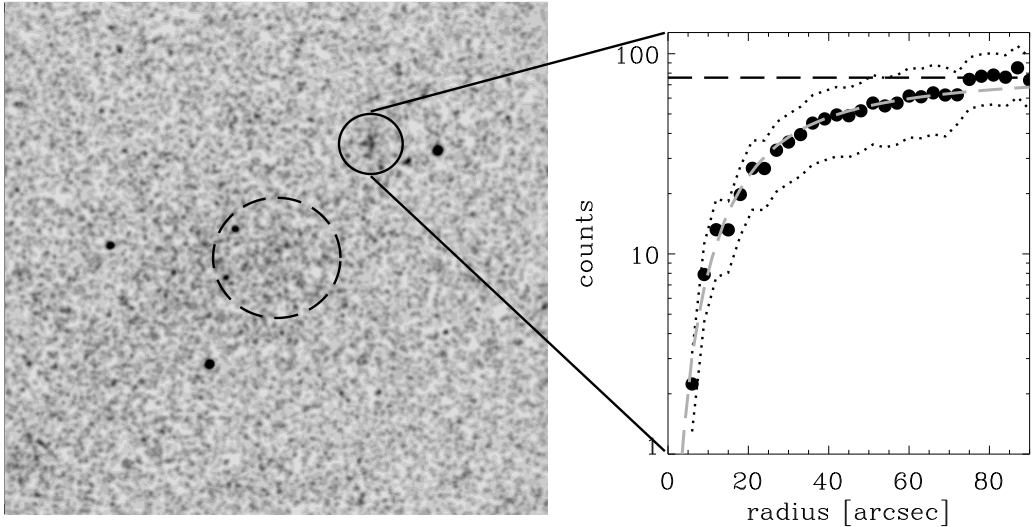}
\caption{ {\bf Left panel:} $24\arcmin \times 24\arcmin$  portion of an HRI
image (Gaussian smoothed with $\sigma=5\, \arcsec$) showing the
detection of cluster candidate BMW044743.4-202042, marked by the solid
90\arcsec~radius circle.  The dashed circle (3\arcmin~radius)
indicates the central area containing the original HRI target (here barely visible),
excluded from the cluster search.  {\bf Right panel:} Integral,
background-subtracted flux growth curve of the candidate cluster
(filled circles), plotted together with its 1$\, \sigma$ error
corridor (dotted lines) and the best-fitting $\beta$-model (Grey
dashed curve).  The total flux of the source is
computed as the asymptotic value of the growth curve (horizontal
dashed line).  From the total flux, the core radius is then derived by
comparison to a grid of $\beta$ profiles convolved with the
instrumental PSF at the specific off-axis angle.
This is a typical source in our catalog, detected at $\sim$ 7\arcmin~off--axis
with 75 net counts in a 10 ks exposure, corresponding to a
flux of $\sim 3\times10^{-13}$ erg  s$^{-1}$ cm$^{-2}$.}  
\label{profile} 
\end{figure*}

The characterization procedure of extended sources has been explained
in C99, but we briefly repeat it here for convenience of the reader.  First, we
considered a sub-sample of sources detected only in the observations
that have a star as target (ROR number beginning with 2). This resulted in 6013
sources detected over 756 HRI fields (Fig.~\ref{sexten}).  The
distribution of source extensions was divided into bins of 1\arcmin, as a
function of the source off-axis angle, and within each bin, a
$\sigma-$clipping algorithm on the source extension was applied: the
mean and standard deviation in each bin are calculated and sources
at more than $3\,\sigma$ above the mean value are
discarded.  In this way we determine the mean instrumental extension
$\left<W_p(\theta)\right>$ and standard deviation $\sigma_p(\theta)$ of
point-like source observed at different $\theta$.  As shown in
Fig.~\ref{sexten}, this provides us with a threshold for defining
truly extended sources (Rosati et al. 1995).   {\it We define as
candidate clusters all sources that lie above the
$3\,\sigma_p(\theta)$ corridor, including their $1\,\sigma_{clu}$
error bar}, i.e. those sources for which, at their observed off-axis
angle $\theta$, one has\footnote{For clarity, note that in the
original BMW--HRI
general catalog, a source was conservatively classified as extended if
its extension W$_{clu}$ and the relative error ($\sigma_{clu}$) lay
at more than $2\,\sigma_{clu}$ from this limit.} 
\begin{equation}
W_{clu}-\sigma_{clu} > \left<W_p(\theta)\right>+3\,\sigma_p(\theta) \,\,\,\,\, .
\label{exte}
\end{equation}
This combined requirement on the distance from the point-like source
locus, and on the intrinsic error in the source extension, roughly
corresponds to a $\sim 3.5\,\sigma$ confidence level for the extension
classification.

\subsection{Significance Criterion}
To reduce the contamination level of the catalogue due to spurious
detections, we also limit the current cluster selection to sources
with detection significance larger than $4\, \sigma$ (see
Sect.~\ref{cont}).  The source significance $p_s$ is defined as the
confidence level at which a source is not a chance background
fluctuation, given the background statistics and the specific field
exposure time.  This quantity
is assessed via the signal to noise ratio in wavelet space. For each
wavelet scale, the noise level is computed through numerical
simulations of blank fields with the corresponding background (L99),
while the signal is the peak of the wavelet coefficients corresponding
to the source. In order to make this significance more easily
comparable to other methods, confidence intervals are expressed in
units of the standard deviation $\sigma$ for which a Gaussian
distributed variable would give an equal probability of spurious
detection (68\%: $1\,\sigma$; 95\%: $2\,\sigma$, etc.). So,
the values of our figure of merit $p_s$ will represent the number
of $\sigma$'s corresponding to the confidence level of that specific
source.  Note that the signal to noise in wavelet space can be different
from that in direct space for several reasons: i) the background
subtraction is more accurate and locally performed; ii) the high
frequencies are suppressed, so that a correlated count excess gives a
higher significance than a random one with the same number of counts;
iii) the exposure map can be incorporated in the wavelet space, so
that artifacts do not affect the source significance.

In the 914 HRI fields considered we have 194 sources meeting these requirements.
Among these, we discovered  that 22 were spurious sources caused
by an hot pixel in the detector (not identified in the data reduction); this, 
coupled with the dithering of the satellite produces a high
signal-to-noise extended source of 2\arcmin~typical dimension always located
at the same detector coordinates.    

We inspected directly each of the remaining 172 sources on DSS2/X--ray overlays 
and cross-correlated their positions with the NED database.
As a result, we removed 18 of them from the catalog as clearly associated 
with a nearby galaxy,
thus ending up with a final list of 154 cluster candidates.
This represents our master statistical sample for follow-up.
Clearly, even at $p_s < 4\, \sigma$ a significant number of sources are
truly clusters.  At this level, however, the contamination rises
to $\sim 40$\% (see Sect.~\ref{cont}), thus making the optical identification of
real clusters significantly less efficient in terms of telescope time.

\subsection{Re-estimation of cluster parameters}
\begin{figure*}  
\begin{tabular}{cc}
\includegraphics[width=8cm]{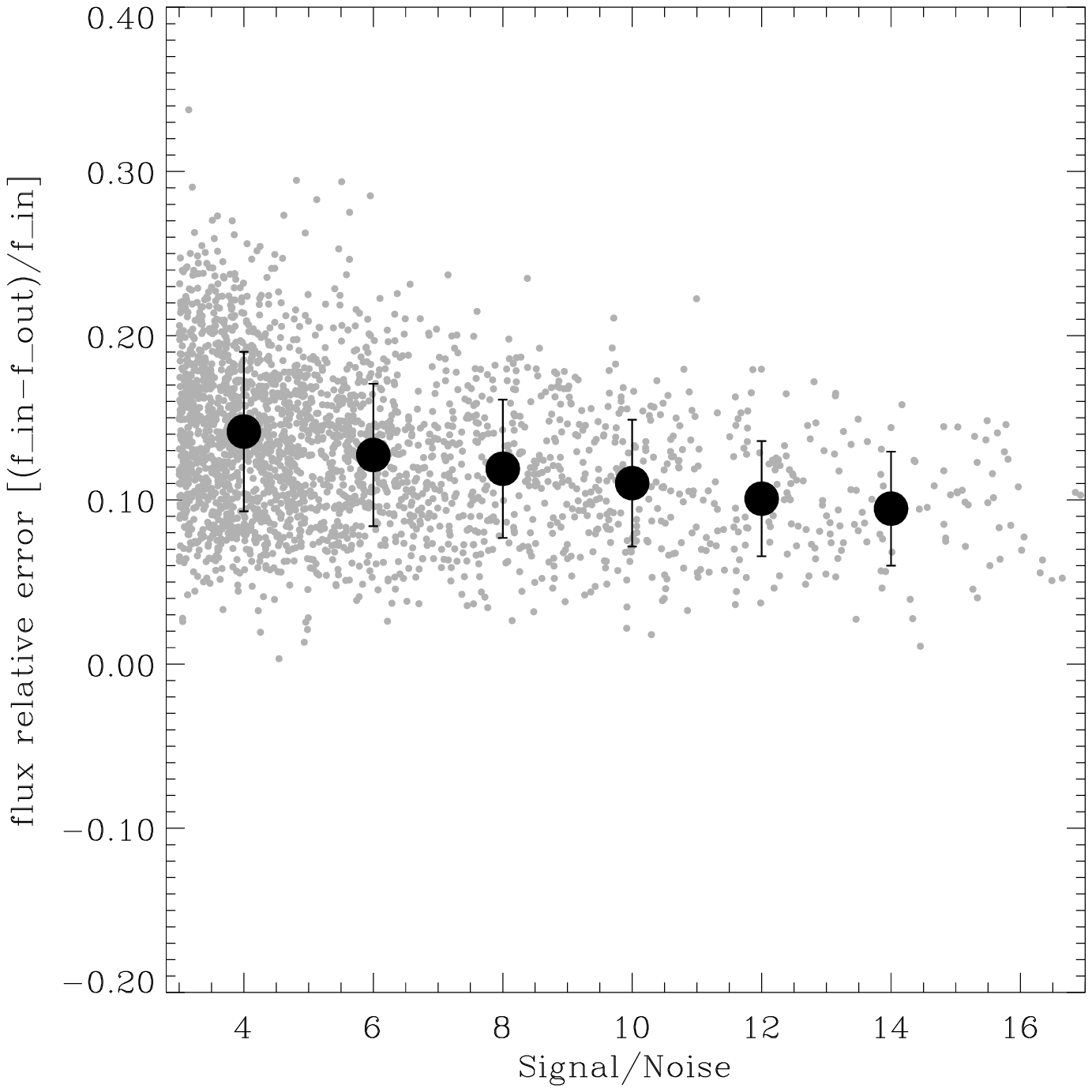} &\includegraphics[width=8cm]{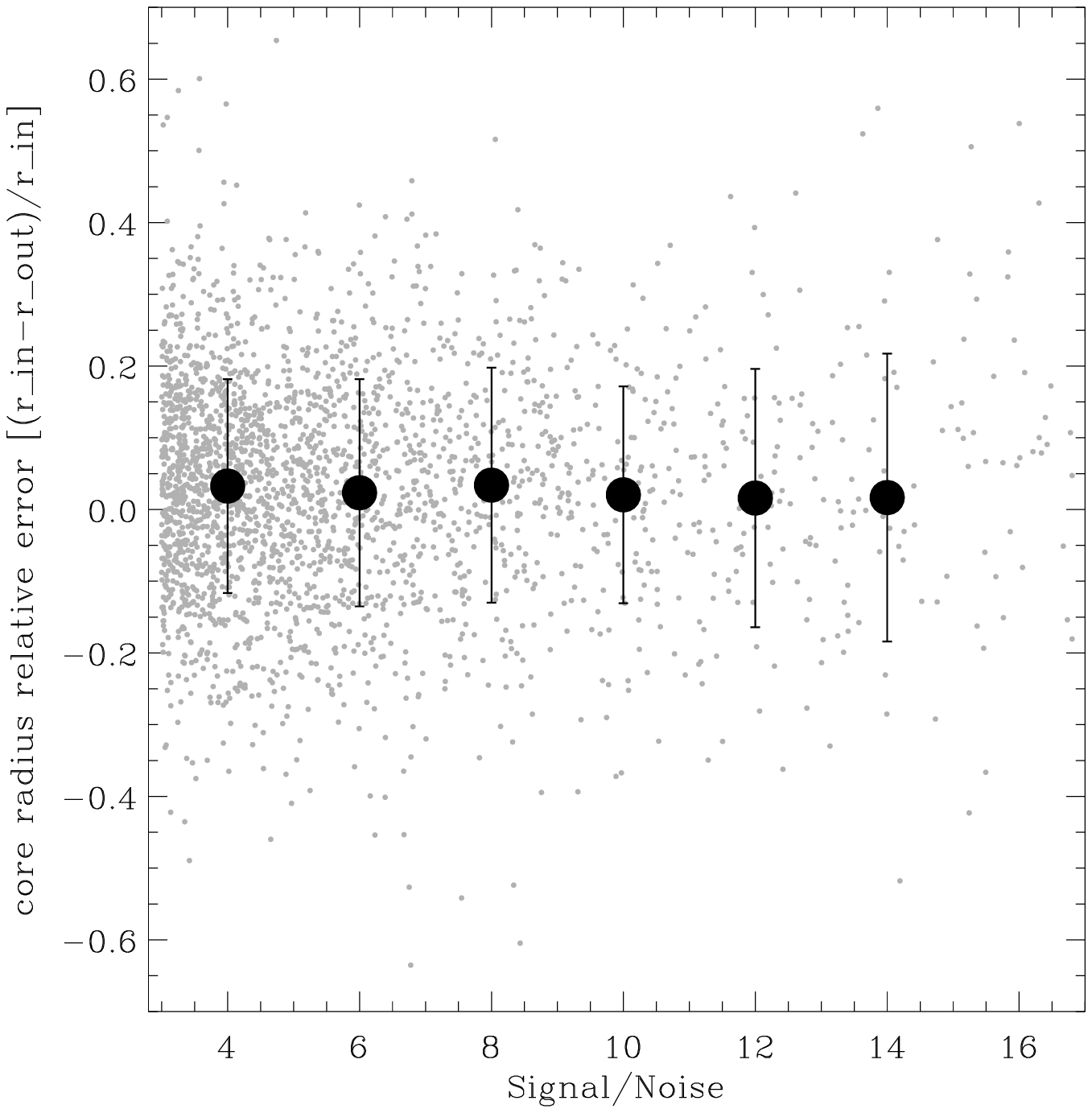}
\end{tabular}
\caption{Relative uncertainties in the flux ({\bf left}) and core radius
({\bf right}) estimation, at different signal-to-noise levels, derived
from the Monte Carlo simulations of the cluster sample. 
They are calculated as the ratio (input-output)/input.
The filled circles give the mean systematic
relative error in each bin and its random standard deviation.} 
\label{inout2} 
\end{figure*}

At the end of the analysis pipeline, the BMW detection algorithm
yields a catalogue with 80 parameters for each source
including the count rate, flux and extension (see P03 and Sect.~\ref{crite}).
These values are reliable for point  sources, but not for sources with a surface
brightness profile (SBP) different from the instrumental PSF, as is
the case for clusters.
The observed cluster profiles are the result of the convolution of the intrinsic SBP with the PSF 
of the instrument. The BMW algorithm calculates the source flux
and extension by means of a $\chi^2$  minimization with respect to a Gaussian 
model in wavelet space (L99, see also previous section). 
Using a more appropriate (non--Gaussian) model within the pipeline is
unpractical.  First, in the general case a cluster model profile
(e.g. a classical $\beta$-model or King profile, see Sect.~\ref{simulations})
convolved with our Gaussian-based wavelet filter (L99) cannot be
handled analytically, so that one of the simplifying features of the analysis pipeline is lost.
Second, a direct profile fit to the data gives reliable results (in
terms of $\chi^2$) only for the few most luminous objects in the
sample, suggesting that an integral approach is the best way to proceed.

For this reason, we adopted a simple and robust ``growth-curve''
technique, as applied to X--ray data by B\"ohringer et al. (2001), 
to re-measure the source total flux. Fig.\ref{profile} visually illustrates
this technique.  The source total flux is defined as the asymptotic value of
the integral SBP, computed within increasing
circular apertures on the background-subtracted images, when this reaches a
horizontal ``plateau''.
The background-subtracted images are obtained from the original images 
by the subtraction of the background map (see Sect.~\ref{simulations}).
Technically, the plateau is defined by studying the local derivative of the 
growth curve, comparing at each step the flux variation within adjacent aperture
radii to the expected mean error.  To eliminate possible contamination from nearby 
sources, we first mask all other sources from the BMW--HRI general catalogue.
Finally, we calculate the conversion from count--rates to fluxes
assuming a {\it bremsstrahlung} spectrum with temperature $T=5$ keV 
 and correcting for Galactic absorption using
the appropriate column density at the source position 
(Dickey \& Lockman 1990).
Once the total flux of the source is established, we estimate its core
radius by fitting its integral profile, adopting for the
differential profile a classical $\beta$--model (Cavaliere $\&$
Fusco-Femiano 1976), with fixed $\beta = 2/3$
\begin{equation}
\label{King}
I(r)= I_0 \left[ 1+ \left( \frac{r}{r_c}  \right)^2 
\right]^{(-3 \beta + \frac{1}{2})} = I_0 \left[ 1+ \left( \frac{r}{r_c}  \right)^2 
\right]^{-{3\over 2}} \,\,\,\,  ,
\end{equation}  
where $r$ in this case describes an azimuthal quantity, with $r_c$ being
the angular {\it core} radius. Choosing a fixed value for $\beta$ is
inevitable, given the low number of photons generally characterizing
our sources, which does not allow a further free fit parameter. In
this way, we obtain an estimate of the source extensions.  This is a
very important quantity, given that, as will be clear in the
following, the statistical properties of the survey strongly depend on the 
extension and surface brightness values of the sources.
As described in the next section, we assess these properties
by means of Monte Carlo simulations: in these tests we coherently
assume for the input simulated clusters the fixed slope $\beta = 2/3$
and variable core radii.

The actual observed SBP is then described by the convolution of the
$\beta$ profile with the instrumental PSF (which depends on the
off--axis position).  In practice, for each cluster candidate
(i.e. total flux), we compute a numerical grid of expected SBPs for
different values of the core radius ranging from 2\arcsec~to
120\arcsec.  We then measure the best fitting core radius by
minimizing the $\chi^2$ between the observed and computed profiles.

The uncertainties in the measurement of these parameters are estimated via a further
specific Monte-Carlo test, whose results are shown in
Fig.~\ref{inout2}.  The left panel of this figure shows that for
sources with S/N$>4$ our flux measurements are systematically
under-estimated by $\sim 12\%$. This is expected, as our fluxes are
integrated out to a certain radius, fixed by the growth-curve plateau.
Ideally, the total flux could be recovered by assuming a model profile
and extrapolating it to infinity. We prefer not to do it, as this
would be model-dependent and would result in larger errors (Vikhlinin
et al. 1998a); rather, we include this uncertainty in the error budget.

Fig.~\ref{inout2} also shows that, as one expects, the flux
measurement errors depend on the source S/N ratio.  We have therefore
divided the cluster sample in S/N bins and assigned to each source the
corresponding relative error from the simulated sample.  One may
wonder why the behaviour of these errors is here studied as a function
of S/N, rather than of the (almost equivalent) $p_s$ confidence level
defined in the previous section.  The reason is that --- unlike source
detection --- the measurement of flux and core radius is performed in
real count space (not in wavelet space).

The right panel of Fig.~\ref{inout2}, on the other hand, gives the
typical errors in the estimates of cluster core radii.  The way the
values of $\beta$ are chosen in the simulations requires some
explanation, as it represents a subtle point.  In fact, if one
constructed input clusters with a fixed $\beta=2/3$, the output of the
simulation would describe the amplitude of statistical errors only
(related to low S/N, etc.).  We know that while $\beta=2/3$
is a very good approximation for most cases, clusters show a
distribution of values around this (e.g. Ettori et al.
2004). This introduces an additional source of error when we try and
measure the cluster core radius with our fixed $\beta$ profile.  In the
simulation, therefore, $\beta$ was distributed according to a Gaussian
with mean value 0.67 and standard deviation 0.05.  With this choice,
the relative error on $r_c$ turns out to be $\sim 20\%$, with
negligible systematic errors as shown in Fig.~\ref{inout2}.

\begin{figure}  
\begin{tabular}{cc}
\includegraphics[width=8cm]{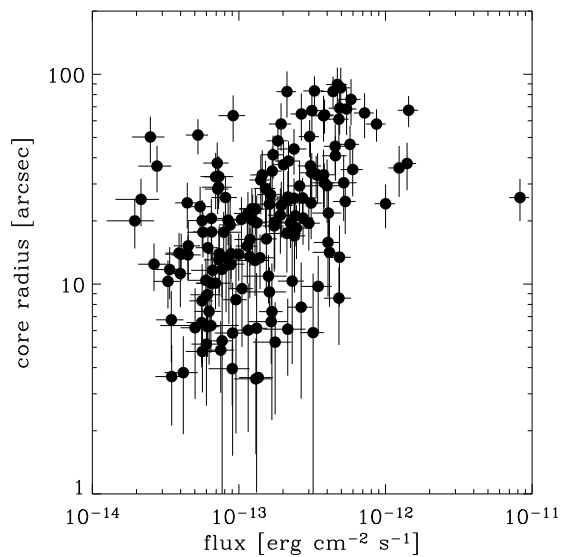} 
\end{tabular}
\caption{Angular sizes (core radii) versus measured X--ray fluxes for the
$p_s>4$ BMW--HRI candidate clusters.    While an intrinsic correlation is
expected between these two quantities, the slope observed here is
probably influenced by the lack of sources at high fluxes, owing to
the small volume sampled.}
\label{rcfl} 
\end{figure}

The measured fluxes and core radii for all 154 sources in our list
are plotted with their errors in Fig. \ref{rcfl}.  The observed
correlation between these two quantities is not surprising if
clusters have an intrinsic distribution of sizes which is not flat
(e.g. Mohr et al. 2001; Arena 2002).  One should consider, however,
that the lack of sources at high fluxes, due to the small survey
area, certainly contributes making the correlation steeper than it
probably is intrinsically.  A discussion of these properties and of
the distribution of angular sizes in the BMW--HRI catalogue goes
beyond the aims of this paper and will be discussed elsewhere (Arena
et al., in preparation).

\begin{figure*}  
\begin{tabular}{cc}
{\includegraphics[width=8cm]{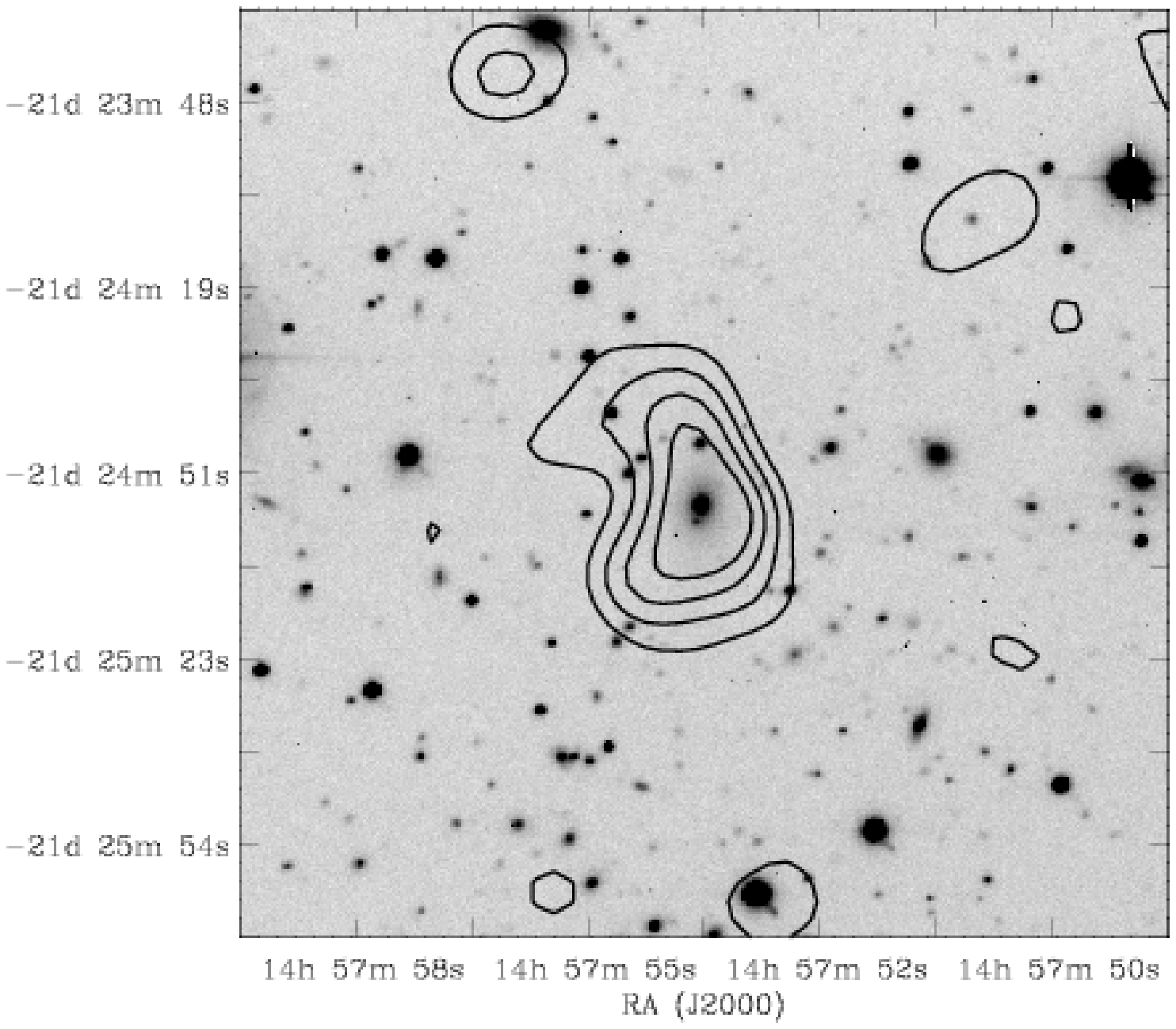}}&{\includegraphics[width=8cm]{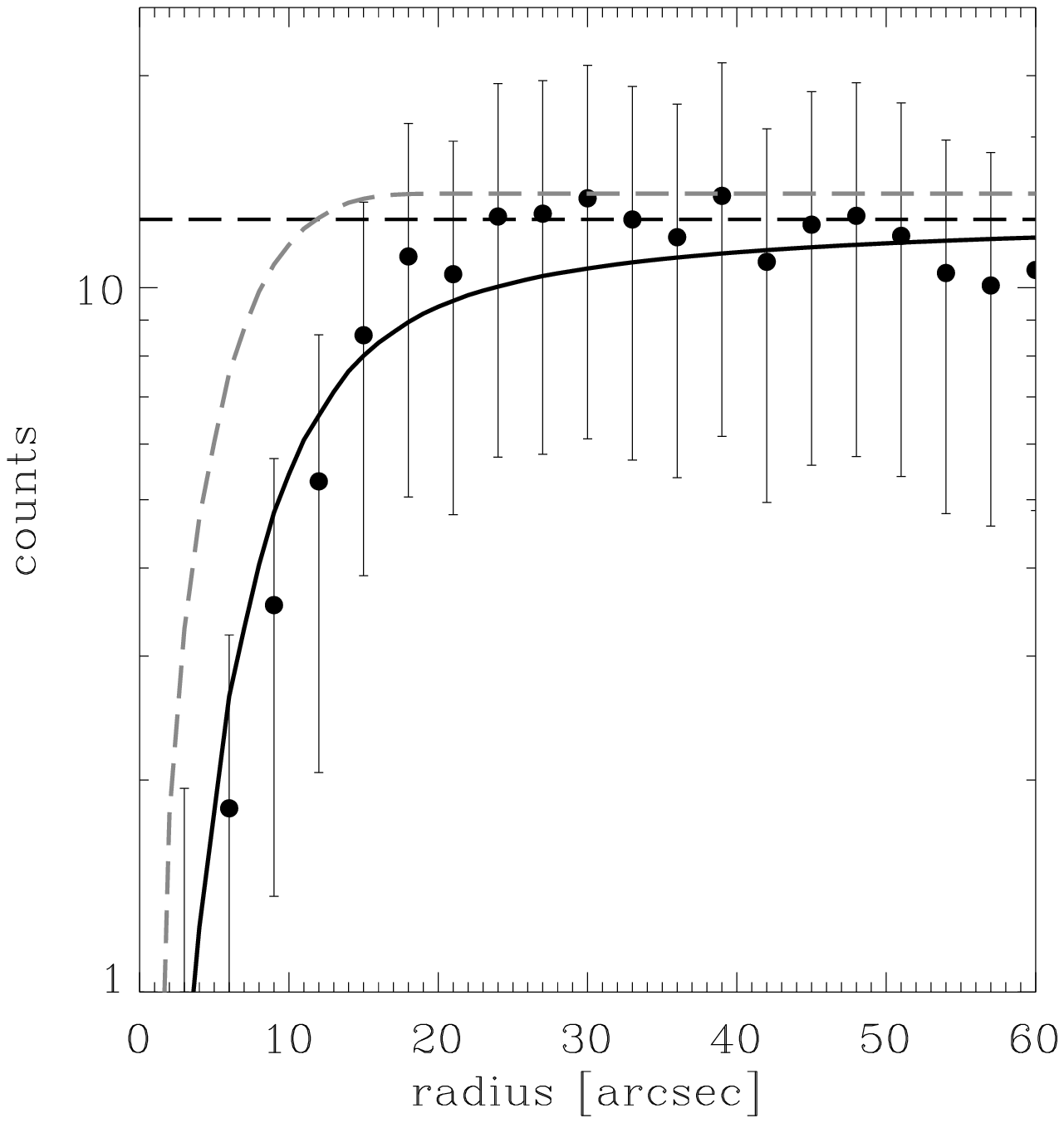}}\\
\end{tabular} 
\caption{A specific example showing one of the major advantages of the BMW--HRI survey,
i.e. its ability to detect as extended (and thus identify as
groups/clusters) also sources with a very small core radius.  This
object, BMW145754.0-212458, has a core radius = 6\arcsec $\pm$ 2 and
would not be detected as extended in a PSPC image.  The
optical follow--up confirmed this source as a real cD--dominated cluster at redshift at $z=0.312$.
In the {\bf left panel} we show a CCD $r-Gunn$ image, with HRI X--ray isophothes overlaid.
The X--ray has been smoothed with a Gaussian filter ($\sigma = 10\arcsec$) and the isophotes
correspond to 1,1.3,1.7,2 standard deviations over the background (after smoothing).   
The {\bf right panel} shows the measured growth curve (filled
circles). The black solid line gives the integral best-fitting PSF-convolved 
$\beta$-model, while the grey dashed line shows the instrumental PSF at the given
off-axis angle.
The horizontal dashed line defines the total flux of the source, as
identified by the growth curve plateau.}
\label{little} 
\end{figure*}

The typical core radius of BMW--HRI cluster candidates measured from this
figure is $\sim$ 20\arcsec.  However, the high resolution of HRI data
allows us to find a number of very small sources, characterized by
core radii as small as $\sim$5\arcsec.  We already know from our CCD
optical follow up that several of these compact sources are not spurious.
Fig.\ref{little} shows a CCD image in the Gunn-$r$ filter (taken with
EFOSC2 at the ESO 3.6~m telescope) and the
corresponding HRI SBP of BMW145754.0-212458, a source with angular
core radius 6\arcsec $\pm$ 2\arcsec.  This candidate has been positively
identified with a cD-dominated group at redshift 0.312 and provides
one key example of the specific advantage of the BMW--HRI catalogue with
respect to previous PSPC-based surveys.  Note that at 
the off-axis angle of this source ($\sim$ 6\arcmin), the instrument
PSF has an half energy diameter of only 8\arcsec.

\section{Monte-Carlo Simulations}
\label{simulations}

To quantify the limitations and assess possible systematic biases in
the detection procedure, we performed extensive Monte Carlo
simulations under realistic conditions.  First, we embedded
simulated clusters and point sources within realistic backgrounds and
ran the detection pipeline, in order to evaluate the {\it completeness} and
the characterization uncertainties of the catalogue.  Second, we ran
the detection procedure on simulated fields containing pure background, 
thus estimating the {\it contamination} of the catalogue due to
spurious detections of background statistical fluctuations.
In this section we describe the simulation results in some detail. 

\subsection{The simulated data}
\begin{figure}  
\includegraphics[width=9cm]{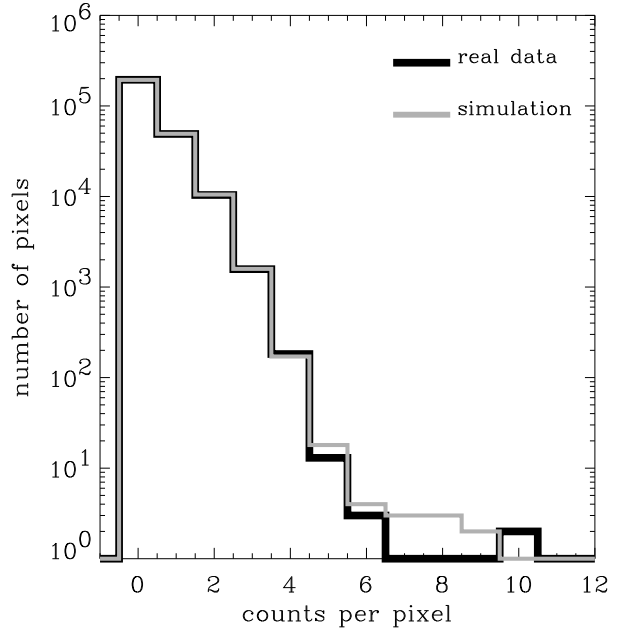}
\caption{Comparison of the global pixel statistics of a
real (black line) and a simulated image (Grey line), for a typical exposure
time of 20,000 sec.  X--ray photons have been binned in pixels of 5\arcsec~
size.  The histograms show how the background statistical properties of the
simulated image accurately match those of the real one. The
simulated background is built as a Poissonian realization of the
background map, and the differences that can be noticed for pixels with
more than five counts are due to the effect of the simulated (extended) sources,
which are overabundant in the simulated image.} 
\label{simpa} 
\end{figure}

\begin{figure}  
\includegraphics[width=9cm]{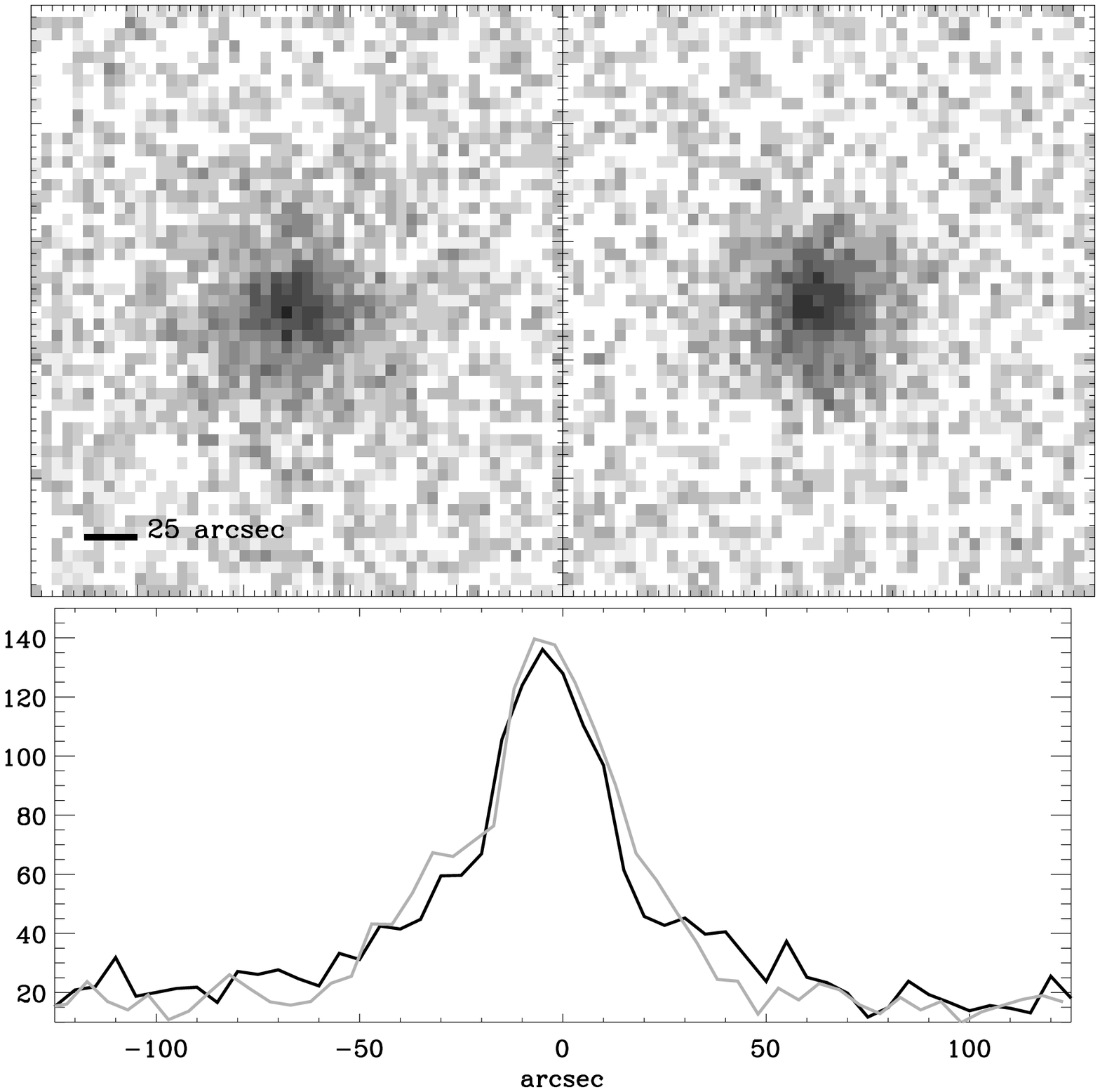}\\
\caption{{\bf Top panel:} visual comparison of the HRI image of
BMW000141.3-154042 (left), the cluster detected with the largest
number of net counts in the BMW--HRI survey,  with a simulated X--ray source
with the same counts and core radius.   The {\bf bottom panel} plots
the average of the central 10 columns of the two images, with the real
cluster described by the black line.
}
\label{true-sim}
\end{figure}

Our simulated sample consists of 12 sets of HRI fields, each set
characterized by a different exposure time evenly sampling (on a
logarithmic scale, see the top-right panel of Fig.~\ref{fita}) the
range of the 914 observations which form the survey.  Specific
predictions for the whole set of actual exposure times are then
obtained by interpolating through these 12 templates.  

For each template field we used the background maps built for the BMW general 
catalog as described in detail in Sect. 3.3 of C99.
 The procedure essentially uses the ESAS software (Snowden 1994),
which produces a vignetted sky background map and a background detector map. 
The total background map is then obtained by summing the detector 
and sky maps. Finally, for each simulated field a Poissonian realization
of such a background is built.
Fig. \ref{simpa} provides an example of how the simulated background
pixel statistics reproduces the observed one in a corresponding HRI
image.

Point--like sources are simulated as pure PSF images with
different normalizations, constructed by means of a ray--tracing
technique (see C99), and their flux distribution is generated so as to
reproduce the observed X--ray number counts in the [0.5-2] keV band
(Moretti et al. 2003).  In this way, we build a fully realistic X--ray
sky, both in terms of background and surface density of point--like
sources, in which to embed the simulated clusters.  To construct
these, we distribute counts according to a $\beta$-model surface
brightness profile (Eq.~\ref{King}), with values of $r_c$ randomly
distributed in the range 5\arcsec-50\arcsec, convolved with 
the PSF at the specific off-axis angle where the cluster is
positioned.  Fig.~\ref{true-sim} shows a visual comparison of a true
cluster in the survey and a simulated one with the same core radius
and integral flux.  In each simulated image, we add {\bf three} such
extended sources, with fluxes randomly chosen in the range
$10^{-14}-10^{-11}$ erg sec$^{-1}$ cm$^{-2}$.  All sources (both
extended and point--like) are located at random positions over the
whole ($\theta=[3\arcmin-15\arcmin]$) field of view.  Overall, in a
typical 20 ksec image this results in a mean number of 15 sources with
more than 10 counts. The rather high number of simulated clusters
allows us to reach a statistically significant number of tests in a
reasonable number of simulations, still avoiding detection biases due
to source confusion problems.  To accumulate a statistically solid
test sample, the procedure is repeated 1000 times for each of the 12
template exposure times.

\subsection{The catalogue completeness}
\label{completeness}
As discussed in the introduction, one of the advantages of X--ray
surveys of clusters is the possibility to properly quantify their {\it
selection function} $S_F$, i.e. the probability that a source with a given
set of properties (e.g. flux, extension) is detected in the survey.
The selection function $S_F$ fully characterizes the sample
completeness and is the key ingredient for estimating its
sky--coverage.

In the ideal case of a uniform observation, given a value $n_{\sigma}$
for the detection threshold in S/N ratio, the
corresponding minimum detectable count rate (or flux limit) can be
expressed by the following formula (see e.g. ROSAT Observer Guide 1992):
\begin{equation}
I_{min}= \frac{n_{\sigma} b^{1/2} d_{cell} }{f_{cell}} \frac{1}{\sqrt{t_{exp}}} + 
\frac{C_{min}}{t_{exp}} \,\,\,\,\,	.
\label{flim}
\end{equation}
Here $d_{cell}$ is the linear dimension of the {\it detection cell} \footnote{Note, as 
an aside, that in our case the wavelet transform,
by its very nature, maximizes the S/N ratio for each source by
optimizing the choice of the detection cell dimension.}, $f_{cell}$ is
the fraction of signal contained in the cell, $b$ is the background
count--rate and $C_{min}$ is the minimum number of source counts
needed for detection when the background is negligible.  However, in
the low S/N regime where the source flux is comparable to the
background noise, the concept of flux limit becomes somehow arbitrary:
sources with fluxes lower than the theoretical sensitivity limit can
be detected with non--null probability.  This is due to the fact that
faint sources can be detected if they sit on a positive background
fluctuation, while they are missed in the opposite case.  For this
reason, the concept of a sharp flux limit needs to be generalized with
a statistical approach, in which the selection function describes the
probability for a source with given properties to be detected.  Given
the characteristics of X--ray telescopes and specifically of the HRI data,
this probability depends in general on: i) the {\it
exposure time of the observation}~; ii) the {\it source position} within the
field of view: moving from the center to the edges of the images, the
X--ray mirror effective area decreases and the spatial resolution
worsens; iii) the {\it source extension}.

Assuming that extended sources can be sufficiently well described by the same
$\beta$-model profile with different core radii, the whole cluster survey can be
characterized by a selection function 
$S_F=S_F(f,\theta,t_{exp}, r_c)$, where $f$ is the flux, $\theta$ is
the off-axis angle, $r_c$ is the apparent extension of the source and
$t_{exp}$ the exposure time of the observation.   Again, the only safe
way to estimate the value of $S_F$ within this multi-parameter space
is by means of Monte Carlo simulations.

\begin{figure*} 
\begin{tabular}{cc}
\includegraphics[width=8cm]{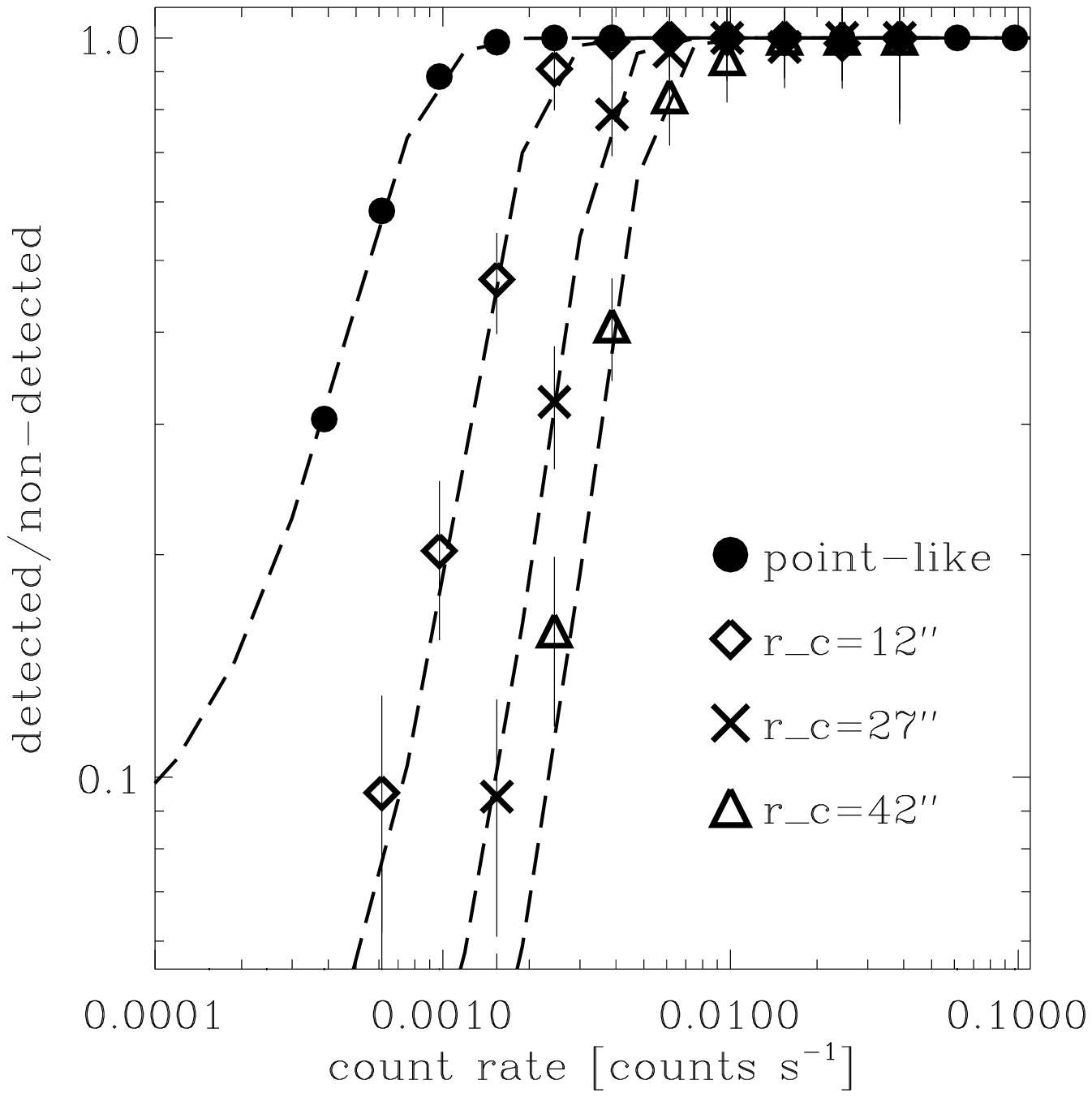} & \includegraphics[width=8cm]{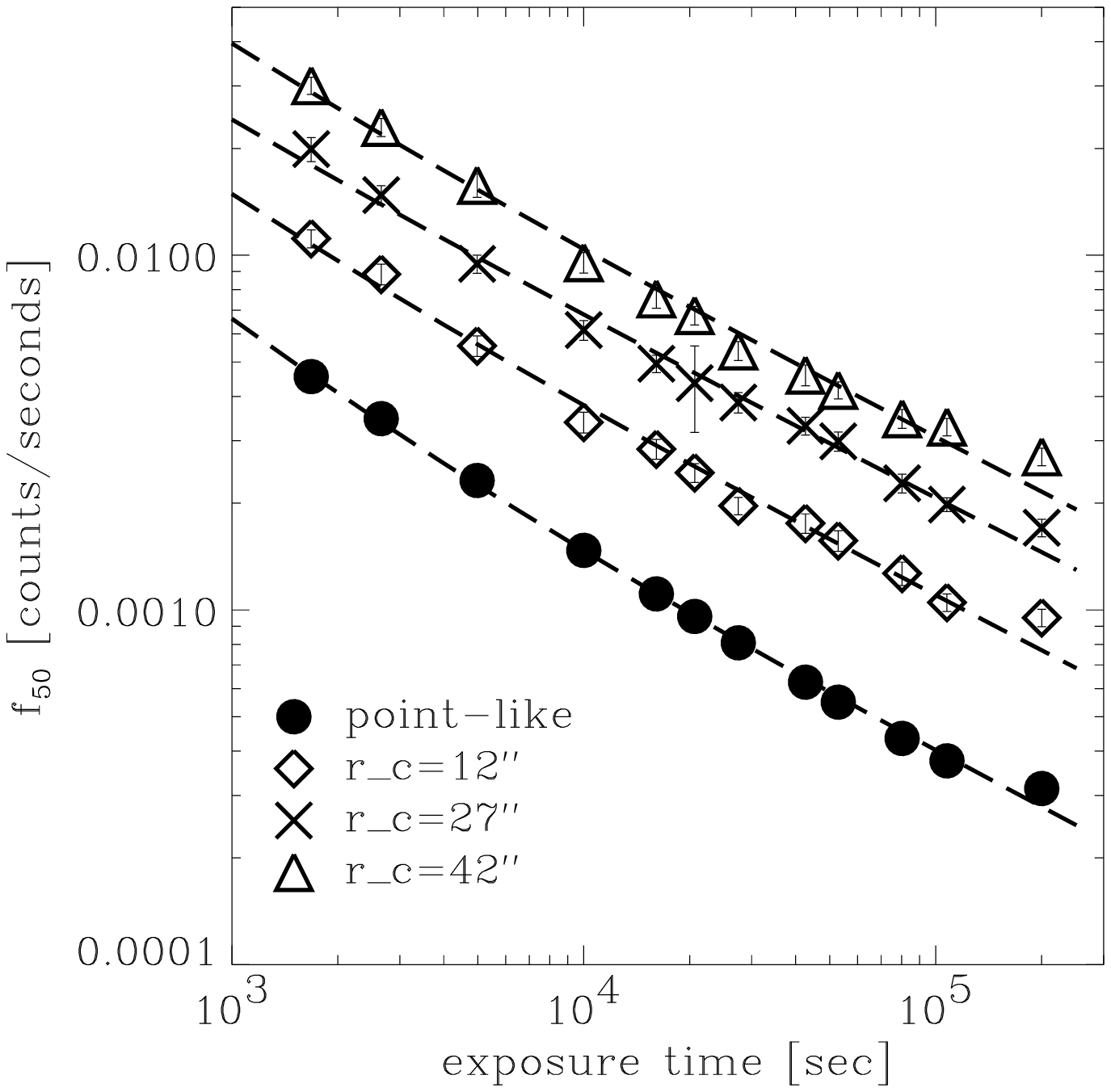} \\
\includegraphics[width=8cm]{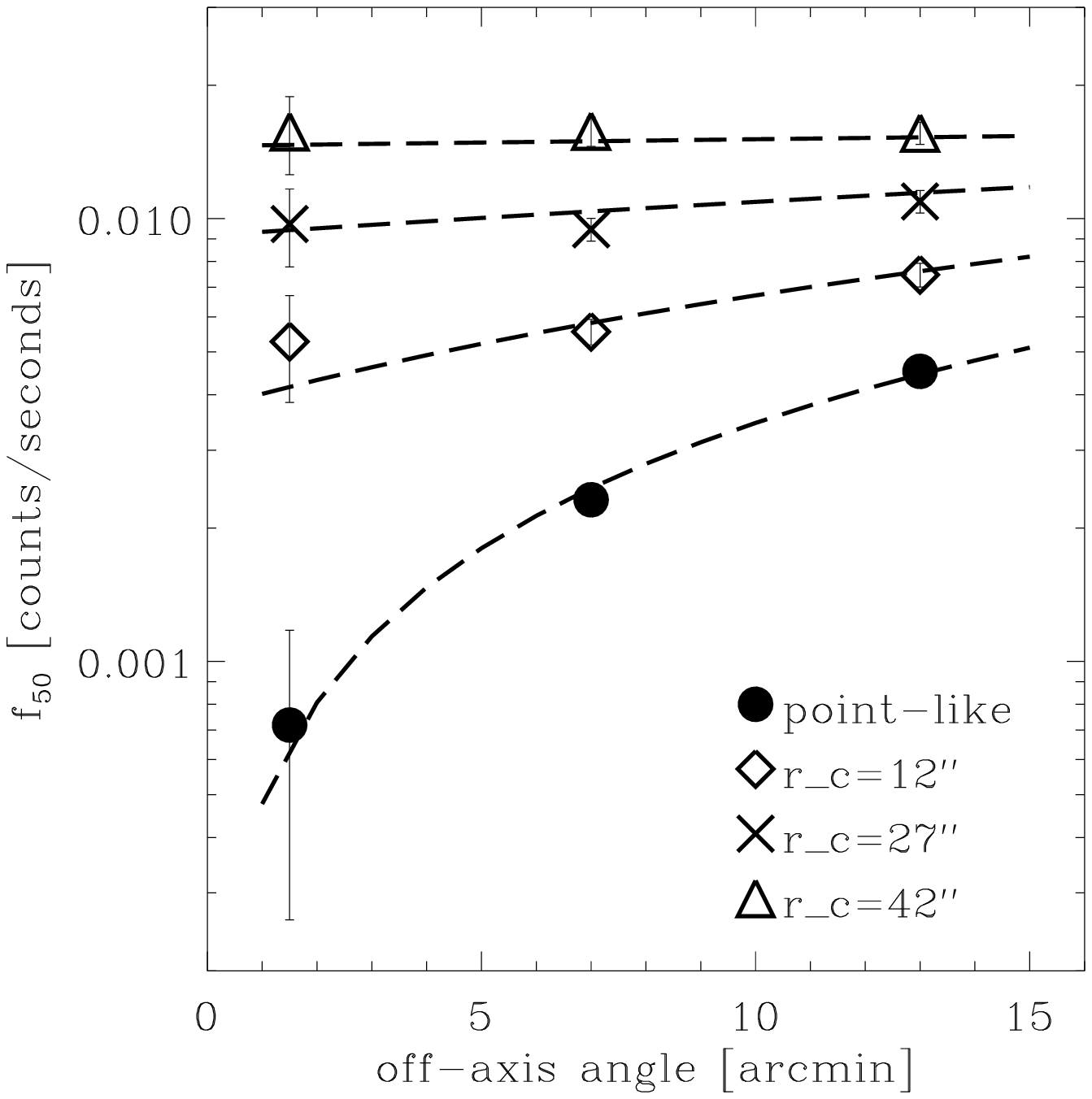} & \includegraphics[width=8cm]{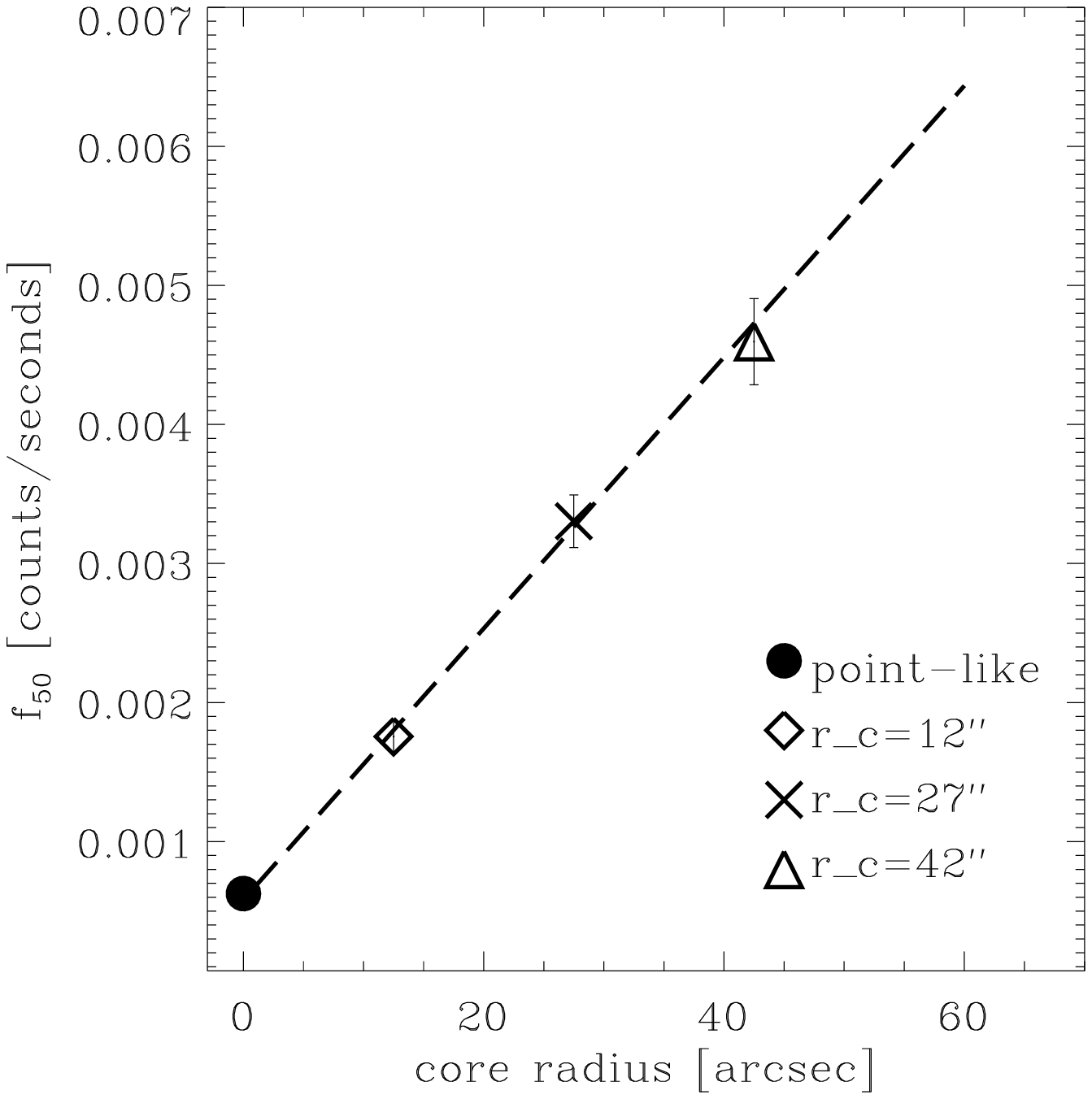}    \\
\end{tabular}
\caption{Dependence of the selection function $S_F$ on the relevant
observational and source parameters, as obtained from the Monte Carlo simulations.  $S_F$ is
defined as the ratio between the number of detected over the
number of simulated input sources, as explained in the text.
In the {\bf top-left} panel we plot the results for the annulus 
between off-axis angle $9\arcmin<\theta< 12\arcmin$, $t_{exp}=40$ ksec
and 4 different SB profiles.  The points are very well fitted by a
Fermi-Dirac function with different values of the two free parameters
(Eq.~\ref{fermif}). 
In the {\bf top-right} panel for all 12 sets of simulated fields, we plot the
values of $f_{50}$ (in counts sec$^{-1}$) for the region $9\arcmin < \theta <
12\arcmin$ as a function of exposure time.  In the {\bf bottom-left}
panel we plot $f_{50}$ as a function of the off-axis angle for a 5
ksec exposure and for four different source extensions.
In the {\bf bottom-right} panel we plot $f_{50}$ as a function of source
core radii, for a 30 ksec observation in the annulus between 3\arcmin~and 10\arcmin.  $f_{50}$ turns out to depend
linearly also on the value of $r_c$.  In all four plots the dashed
lines reproduce the multidimensional fit as explained in the text.}
\label{fita} 
\end{figure*}

To this end, we first grouped the input simulated sources into a grid
defined by 3 different angular extension ranges, 
$r_c=[5\arcsec-20\arcsec]$, $r_c=[20\arcsec-35\arcsec]$,
$r_c=[35\arcsec-50\arcsec]$, together with 20 logarithmic bins in count rate.  We
then considered in each of the 12 simulated sets of images, 3
different radial areas, defined by the annuli between
$[3\arcmin-9\arcmin]$, $[9\arcmin-12\arcmin]$, and $[12\arcmin-15\arcmin]$
off-axis angles.  At this point, we could compute the actual values of
the selection function $S_F$ by measuring the ratio of the number of
detected to the number of input simulated sources within each bin of
the 4-dimensional hyper-space defined by exposure time $t_{exp}$, flux
$f$, core radius $r_c$ and off-axis angle $\theta$.  The four panels
in Fig.~\ref{fita} summarize the results of this procedure.  The top-left
panel considers a simulated field with $t_{exp}=40$ ksec and the
annulus between 9\arcmin~and 12\arcmin.  The points and curves
show how the ``projected'' $S_F$ for different source
extensions and as a function of flux is well fitted by a Fermi-Dirac
function: 
\begin{equation}
S_F(f,\theta,r_c) = \frac{1}{e^{\frac{f_{50}-f}{c}}+1} 
\label{fermif}	
\end{equation}
\noindent 
where $f_{50}$ and $c$ are the two free parameters of the functional
form.  The parameter $f_{50}$ corresponds to the flux where $S_F$ equals
0.5, or in other words, the detection probability equals $50 \%$, while
$c$ simply describes how sharp the cut-off in the selection function is.

Considering different exposure times and off-axis angles, one gets 
analogous plots, with different values of the pair $(f_{50},c)$.  By
constructing a 3-dimensional grid of the best fit $(f_{50},c)$ values
over (a) 12 different exposure times, (b) 3 off--axis angles and (c) 4
core radii, we then obtain the general expression of the
selection function $S_F$ over the whole parameter space
(see panels in Fig.~\ref{fita}).  In this way, for a source with a given core
radius within the range 5\arcsec-50\arcsec, we have the detection
probability for an arbitrary exposure time observation (ranging from
1 ksec to 200 ksec ), at an arbitrary position within the field of view
(3\arcmin $<~ \theta <$ 15\arcmin) and for an arbitrary flux (ranging
from $ 10^{-15}$ to $10^{-11}$ erg sec$^{-1}$ cm$^{-2}$).  For the few
clusters with core radius larger than 50\arcsec, $S_F$ is safely
obtained by linear extrapolation, according to the bottom-right panel
of Fig.~\ref{fita}.

\subsection{The catalogue contamination}
\label{cont}
By construction, the detection threshold in the overall BMW HRI source
catalogue is set by fixing the expected number of spurious sources:
the higher the number of allowed spurious sources, the lower the
threshold.  This number is set to {\bf 1 in ten fields} for each scale
of the wavelet transform, which corresponds to 0.4 expected spurious
sources in each frame. This, however, should be taken as a
theoretical estimate because it assumes a perfectly flat background so as
to provide a general instrument-independent reference (L99).  Here, we
derive a more accurate estimate of the number of spurious detections
by properly simulating a realistic background. Moreover, due to the
empirical definition of extended sources, we cannot analytically
predict the number of spurious sources that will be detected as
extended and need to resort again to a Monte Carlo test.  To this end,
we build pure background images for the usual set of 12 templates with
different exposure times (and different background values) ranging
from 1 ksec to 200 ksec (see Sect.~\ref{simulations}).
This time, we do not add any simulated source.
For each of the 12 template images, we generate 200 different
realizations of the background map, assuming a pure Poissonian noise.  
Then we run the BMW detection pipeline: clearly, all detected sources will be
spurious, resulting from background fluctuations above the detection
threshold.  

Within the 2400 simulated blank frames, we detect 844 spurious sources (both point--like
and extended), to be compared with the expected value of $0.4 \times
2400 = 960$.  The close agreement between these two figures represents
an encouraging {\it a posteriori} confirmation that the total number
of spurious detections is well under control in the whole selection
procedure.   Among these, 108 sources meet the requirements of our 
cluster survey (extension and significance), providing us with an
estimate of 0.045$\pm 0.003$ spurious $p_s>4\,\sigma$ detections in each frame
of the survey. Note that this value does not depend on the specific
exposure time of the field, as in each field the detection threshold is pushed
down only to the appropriate level so as to keep the total number of
spurious detections constant.  Given the 914 fields composing our
survey, we thus expect 38$\pm 3$ spurious sources to contaminate the
sample of 154 cluster candidates (25$\%$).  Note that the expected number of
spurious detections is proportional to the number of the analyzed
fields, not related to the number of sources in the catalog.  This
means that, for example, if one concentrated only on a subset of
high-exposure fields, the number of spurious sources would go down
faster than the number of true clusters.  However, the explored area
and volume would also be reduced accordingly, thus affecting the ability
to find luminous, rare clusters, which is one of our goals.

There is a second source of contamination that needs to be considered,
which is due to point-like sources mistaken as extended.   This
fraction is readily estimated from our first set of simulations, where
we realistically simulated the observed sky in terms of surface
density distribution of point--like sources.   The net result is that
in the sample of 154 $p_s>4\,\sigma$ cluster candidates, we expect 
4$\pm 2$ sources to be false classifications of a point--like source,
which makes up for a total contamination of $42\pm 4$, i.e. 27$\%$.

\begin{figure}  
\includegraphics[width=8cm]{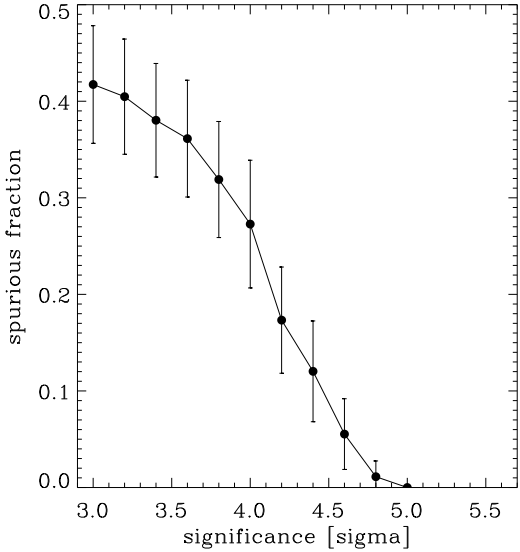}
\caption{The expected number of spurious detections in the BMW--HRI
cluster catalogue as a function of the source significance, expressed in units
of standard deviations $\sigma$ (see text for details).  The curve
includes the contamination produced both by background fluctuations
detected as true (extended) sources and by real point sources mistaken as
extended by the algorithm.   On the basis of this curve, the BMW--HRI
$p_s>4$ master sample is expected to have a 27\% contamination,
while selecting at $p_s>5$ one obtains a sample
with virtually null contamination.}
\label{contam} 
\end{figure}

As shown by Fig.~\ref{contam}, the number of total expected
spurious detections (background fluctuations plus wrongly classified
point-like sources), decreases significantly at higher detection
significance: it is more than $40\%$ at $3\,\sigma$, $27\%$ at $4\,\sigma$
and negligible beyond $5\,\sigma$.   The chosen significance
threshold at $4\,\sigma$ for our main cluster catalogue is based on
this plot, representing a good compromise between the total number of
sources and the expected contamination.   

\section{Sky Coverage and Number Counts}

\subsection{The BMW--HRI Cluster Survey Sky--coverage}
\label{skycov}
\begin{figure} 
\includegraphics[width=8cm]{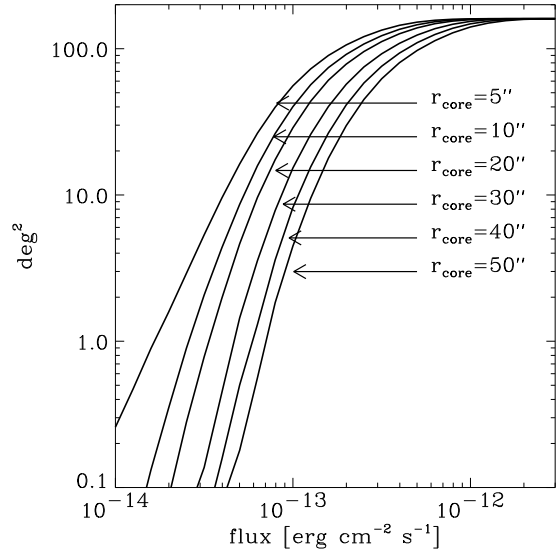}
\caption{The solid angle covered by the BMW--HRI survey (sky--coverage) as
a function of flux and source extension, obtained by integrating the
survey selection function over the observed fields.}
\label{skyc} 
\end{figure}

The sky--coverage $\Omega$ (SC hereafter), measures the actual surveyed area of the
survey as a function of X--ray flux and of the other parameters characterizing
each source: it is a necessary ingredient for any statistical computation
involving the mean surface density of clusters in the catalogue, like e.g.
the log N --log S or the luminosity function.   

By definition, it is closely related to the survey selection function,
which, as we have seen, depends also on the off-axis angle $\theta$ and
the source size $r_c$ (see Sect.~\ref{cont}).  Let us therefore consider
an infinitesimal annulus at an off-axis angle $\theta$, characterized
by a solid angle $d\omega$, within the field of view of a generic
observation $i$.  The area of this elementary surface will contribute
differently to the total sky coverage, depending on the flux limit one
chooses and on the source size: for example, depending on the
exposure time of the observation, it will provide its full area at a
given bright flux, while it will be ``invisible'' at fainter fluxes.
Or, equivalently, it will be effective for finding very small sources,
and become ineffective for very large, blurred sources.  All these
effects are already taken into account by the selection function $S_F$ that we
have carefully estimated through our Monte Carlo simulations.  
In fact, by definition, the actual
contribution of this annulus to the total sky coverage will be
\begin{equation}
d\Omega_i=d\Omega_i(f,\theta,r_c)=S_{F_i}(f,\theta,r_c)d\omega  \,\,\,\,\, ,
\end{equation}
and the total SC yielded by that given field $i$ for a source with
flux $f$ and size $r_c$ will then be simply the sum over the annuli, i.e.
\begin{equation}
\Omega_i(f,r_c) = \int_{\theta_{min}}^{\theta_{max}} S_{F_i}(f,\theta,r_c) d\omega(\theta)\,\,\,\,\, ,
\label{ome_uno}
\end{equation}
where in our case $\theta_{min}=3\arcmin$ and
$\theta_{max}=15\arcmin$. 
Finally, the overall SC of the survey, $\Omega_{BMW}(f,r_c)$ will be given by
the sum of the  contributions by each single observation:
\begin{equation}
\Omega_{BMW}(f,r_c) = \sum_i \Omega_i = \sum_i
\int_{\theta_{min}}^{\theta_{max}} S_F(f,\theta,r_c) d\omega(\theta) \,\,\,\,\, ,
\label{ome_tot}
\end{equation}

which shows explicitly the dependence of the total sky coverage on the
source extension $r_c$ (Fig.\ref{skyc}).  Thus, given an observed
source characterized by an $(f,r_c)$ pair, it will always be possible to
give a value for the total solid angle over which that source could
have been found in the survey.  

\begin{figure} 
\includegraphics[width=6.5cm,angle=270]{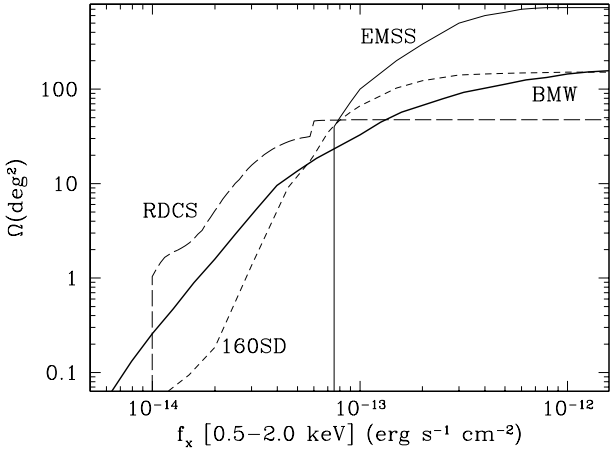}
\caption{Comparison of the BMW--HRI size-weighted sky coverage to those of
the EMSS (Gioia et al. 1990), 160 Square Degrees (Vikhlinin et
al. 1998a) and RDCS (Rosati et al. 1998).  Since the sky coverage is a
function of both flux and source extension, the BMW--HRI curve has been
computed combining the sky coverages pertaining to different
source extensions, according to the observed distribution of angular
sizes in the $p_s>4$ sample.}
\label{skysurv} 
\end{figure}

Given the various dependences we have discussed, a comparison of the sky
coverages as a function of X--ray flux among
different surveys can only be done in an approximate way,
i.e. considering a ``typical'' sky coverage for each specific survey
(e.g. Rosati et al. 2002).    To perform such a comparison, given that
the source extensions correlate with their fluxes (Fig.~\ref{rcfl}), 
we tried and obtain a sufficiently realistic one-dimensional $\Omega(f)$ by
computing the mean $r_c$ values within 5 bins in flux, and then
constructing a ``composite'' 1D sky coverage which in each flux
range reflects the typical size of sources with that mean flux.
The result is shown in Fig.~\ref{skysurv}, compared to the sky
coverages of three other representative surveys from the literature.
This plots shows that the BMW--HRI survey is competitive with existing PSPC
surveys both at  bright and faint fluxes.  In particular, it provides
an interesting solid angle coverage below 6$\times$10$^{-13}$ erg sec$^{-1}$
cm$^{-2}$, which results in the potential ability to detect a few
clusters beyond $z\sim 0.9-1$.  In fact, among the few BMW--HRI clusters
with confirmed spectroscopic redshift, we already have two
objects at $z=0.89$ and $z=0.92$.    We remark that these values are specific for the
low-contamination $p_s>4\,\sigma$ sample.  In Sect.~\ref{sdndz} we shall use this 1D sky coverage 
to compute the expected redshift distribution of BMW--HRI clusters.

\subsection{Accounting for overlapping fields}
As anticipated in Sect.~\ref{multi-area}, $7 \%$ of the survey area was
observed more than once. Therefore, when summing the areas contributed by
each field in the sky coverage calculation we need to
account for this repeated area so as to include it only once.
To this end, let us
consider a given source with flux $\bar f$, which has been observed $N$
times with different exposure times and therefore with different
values of the selection function.  In other words, the {\it
probability $P_i$ of detecting this source}  is different in each exposure
$i$, coinciding with the value $S_{F_i}$of the $i$-th selection
function at flux $\bar f$. Accordingly, the {\it global} selection function for that
specific source $S_F(\bar f,r_c)$ in the N observations will correspond to the probability of
detecting the source within {\it at least}  one of the $N$ observations.
If we consider the probability $Q$ of {\it not detecting} it in {\it
any} of the observations, this is nothing else than the product of the
probabilities of not detecting it in each image
\begin{equation}
Q=(1-P_1)\times(1-P_2)\times...(1-P_N) = \prod_i (1-S_{F_i})\,\,\,\,\, .
\end{equation}
At this point, the selection function pertaining to that source,
$S_F(f,r_c)$ will be just the complement of $Q$, i.e. 
\begin{equation}
S_F=1-\prod_i (1-S_{F_i})\,\,\,\,\, .
\end{equation}

\subsection{The expected redshift distribution}
\label{sdndz}
Using the computed sky coverage, we can now address one of the most
interesting aspects of a deep cluster survey, i.e. the expected redshift
distribution of the sample.   In particular, it is of interest to ask
how many clusters we expect in the cosmologically interesting range
$z\sim 1$, if any, in different evolutionary scenarios.

\begin{figure*} 
\begin{tabular}{cc}
\includegraphics[width=8cm]{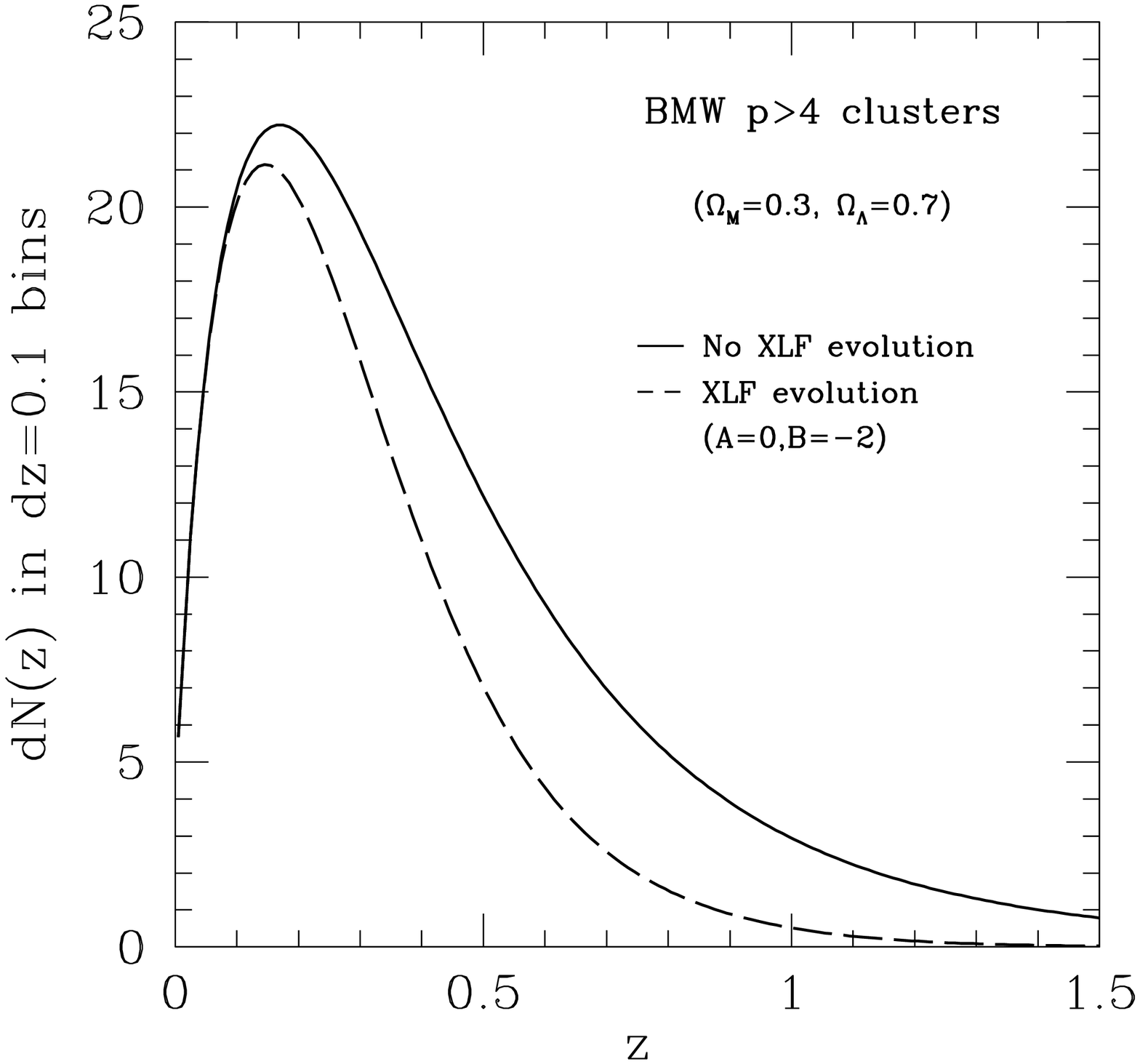}&\includegraphics[width=8cm]{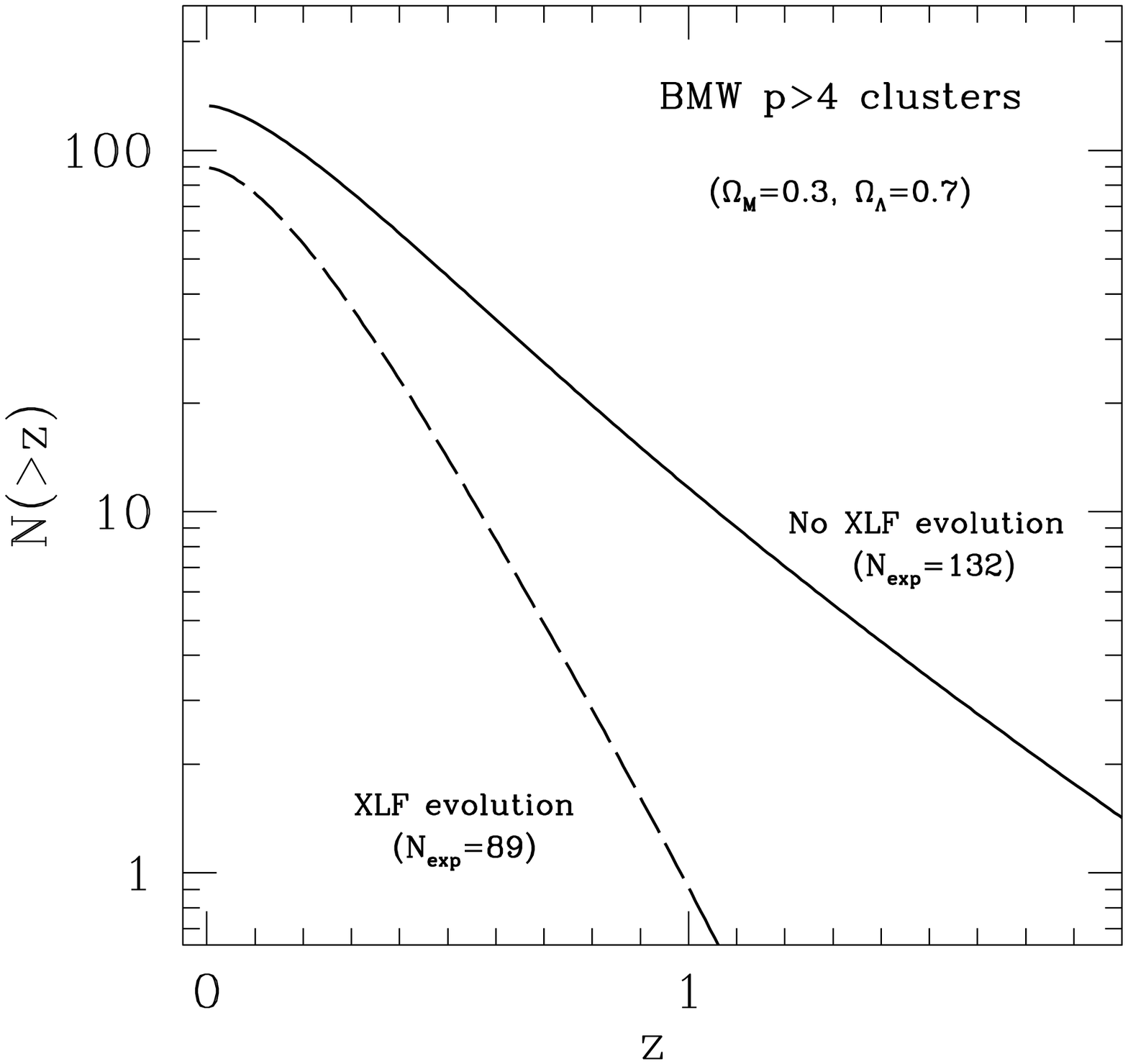}\\
\end{tabular} 
\caption{The expected redshift distribution of the BMW--HRI cluster sample,
in differential form ({\bf left}), showing the number of clusters
expected in bins of $\Delta z=0.1$, and in integral form ({\bf
right}), showing the number of clusters expected above a given
redshift $z$.  The dashed curves in both panels refer to an empirical
evolution model for the XLF as defined by Rosati et al. (2000), using
a conservative pair of parameters {\it(A=0,B=-2.5)}, which best
describe the overall behaviour of available PSPC surveys (Mullis et
al. 2004).
}
\label{dndz} 
\end{figure*}

Briefly, the differential redshift distribution is obtained in the
standard way as
\be
{dn\over dz} = {dn\over dV} {dV\over dz} = {dn\over dV} {dV\over dl}
{dl \over dz} \,\,\,\, ,
\ee
where $dl/dz$ is the cosmological comoving line element, $dV/dl =
d_A^2 \,d\omega$, $d_A$ is the angular size distance and $d\omega$ is
the elementary solid angle covered (e.g. Peebles 1993, pag.~331).  The
sky coverage enters here when we integrate over the whole observed
solid angle, while $dn/dV$, i.e. the number of clusters per unit
volume expected at the same $z$, is obtained by integrating the XLF from
the minimum allowed luminosity (at that $z$) to infinity,
\begin{equation}
 {dn\over dV}(z) = \int_{L_{min}(z)}^\infty {dn\over {dL_X\,dV}} dL_X = \int_{L_{min}}^\infty \phi(L_X) dL_X\,\,\,\, ,
\label{xlf}
\end{equation}
where $\phi(L_X)$ is very well described by the usual Schechter functional form, yielding
\be
{dn\over dV}(z) = \int_{L_{min}}^\infty \phistar \left({L_X\over \lstar}\right)^{-\alpha} 
 \exp{(-L_X/\lstar)}\, {dL_X \over \lstar}\,\,\,\, .
\ee
We therefore used the XLF parameters for the [0.5-2] keV band measured
by the REFLEX survey (B\"ohringer et al. 2002), re-computed for a
lambda-cosmology by Mullis et al. (2004), $\phistar=8.56\times
10^{-7}\, {\rm h}^{3}$ Mpc$^{-3}$, $\lstar=1.295\, {\rm h}^{-2}$ erg
sec$^{-1}$, $\alpha=-1.69$, to obtain the solid curves shown in
Fig.~\ref{dndz}, under the hypothesis of a non-evolving XLF.
Rosati et al. (2000) first used a Maximum-Likelihood approach to
quantify in a phenomenological way the evolution observed in the XLF
measured from the RDCS\footnote{A similar ML approach
was used in a more physical way to estimate the values of $\Omega_M$ and
$\sigma_8$ from the observed evolution (Borgani et al. 2001)}.  To 
this end, they parameterized evolution in density and luminosity
through a simple power-law model, in which 
\begin{equation}
\phistar(z) = \phistar_0\,(1+z)^A\,\,\,\, ,
\label{ABmodel1}
\end{equation}
\begin{equation}
\lstar(z) = \lstar{_0}\,(1+z)^B\,\,\,\, 
\label{ABmodel2}
\end{equation}
and $\phistar_0$, $\lstar{_0}$ are the XLF parameter values at the current
epoch.  From their analysis of the RDCS survey they estimated
$A=-1.2$, $B=-2$, indicating a statistically significant evolution in
the mean density of luminous clusters.  The same model has
recently been applied to the 160SD survey data by Mullis et al. (2004),
where also the EMSS, NEP and RDCS constraints are added. By comparing
all these available estimates, they suggests
that the best fitting values for the parameters A and B are $A\sim 0$,
$B\sim -2.5$.  These are the values we adopted for our computations of
the expected number of clusters in the BMW--HRI survey with evolution of
the XLF.

The overall results, with and without evolution, are compared in
Fig.~\ref{dndz}.  The first interesting point to notice concerns the
total number of clusters predicted in the two scenarios.  We have a
prediction of a total of 89 to 132 clusters in the $p_s>4$ sample,
depending on evolution.  Our current number of candidates is 154, with
a prediction of 42 being spurious due mostly to background
fluctuations.  This yields a number of 112 true clusters
in the survey, which is right between the predictions of the two
curves.  This is interesting, as it could imply that less evolution of
the XLF is seen in the BMW--HRI sample.  However, note that the simple
{\it (A,B) }evolutionary model has been applied in the original form
(Borgani et al. 2001), which, taken at face value, while reproducing
correctly the observed deficit of luminous clusters above $z\sim 0.6$,
also under-predicts the number of intermediate-redshift systems, as can be
seen from the left panel of Fig.~\ref{dndz}, where the two curves already begin to separate at
at redshifts as low as $z=0.15$, where we
know that no evolution whatsoever is observed.  One possibility is to
change slightly the Borgani et al. functional form describing the
evolution of the XLF, as done by Mullis et al. (2004), by introducing
an effective non-zero reference redshift in Eq.~(\ref{ABmodel1}) and
(\ref{ABmodel2}) which shifts the ``switch-on'' of evolution to larger,
more realistic redshifts.

The second interesting point to be appreciated from the figure is the
significant number of high redshift systems that the BMW--HRI survey should
be capable of detecting.   Even with the strongest evolution of the
XLF we expect 3 clusters with $z>0.8$, with the number rising quickly
to $\sim 10$ if milder evolution is admitted.   In fact, we already have
three confirmed clusters in this range (see Sect.~\ref{deep}),
with a handful of further confirmed candidates with photometric
redshift $>0.7$.   One of these is an extremely promising object
showing a concentration of galaxies with $r-Ks=[5.5-6]$ around the
X--ray source position, with a range of colours typical of early-type
galaxies at $z\sim 1.2-1.3$ (see e.g. Stanford 1997).
    
\subsection{The $Log N -- Log S $}

The cluster number counts (historically known also as the
``$Log N -- Log S $'' distribution) are the simplest diagnostic of
cosmology/evolution that can be obtained from a sample of cosmological
objects, without knowing their distances.  It also represents a basic test for
the quality of the sample and/or the reliability of the sky coverage,
being sensitive to residual biases and contaminations as a function of flux.  
\begin{figure*}  
\includegraphics[width=15cm]{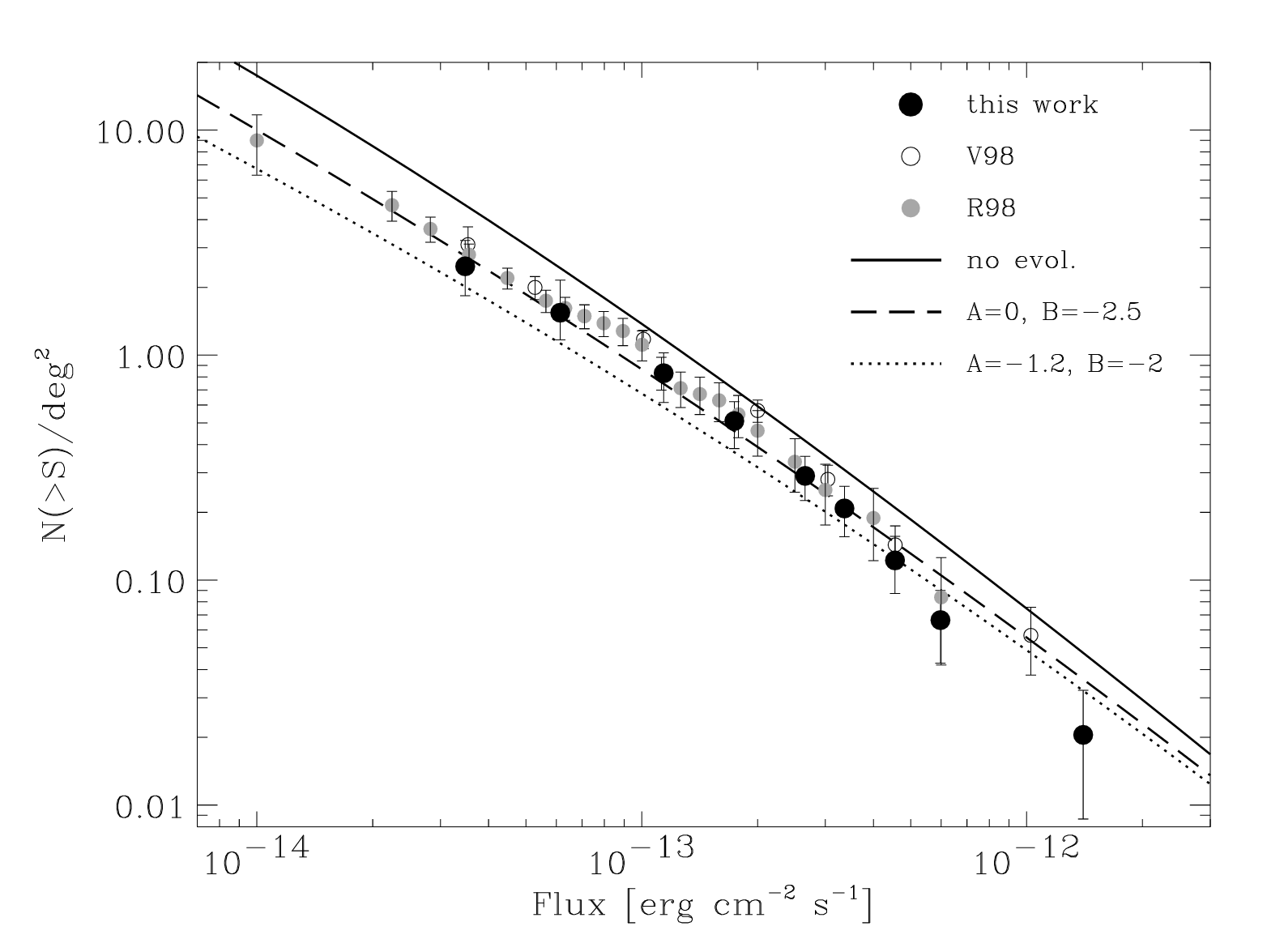}
\caption{The classical $Log N -- Log S $ plot --- i.e. the mean cumulative surface
density of clusters as a function of X--ray flux --- from the BMW--HRI
survey (filled circles), compared to the RDCS (Rosati et al. 1998) and
160SD survey (Vikhlinin et al. 1998b).  The BMW--HRI points are 
computed from the high-significance $p_s>5\, \sigma$ sub-sample, for
which the expected contamination is negligible. The curves give the
expected counts in the same $\Lambda$-cosmology adopted in this paper,
for the cases of no-evolution and evolution of the XLF. The BMW-HRI
and PSPC data are all consistent with a negligible to moderate
evolution of the XLF.
 }  
\label{lnls} 
\end{figure*}

To compute the integral flux distribution, we weight each source by
the inverse of the total area in the survey over which the source
could have been detected.  This is nothing else than the inverse of
the sky--coverage value at the specific source flux and extension:
\begin{equation}
n(>S) = \sum_{f>S} \frac{1}{\Omega_f}
\,\,\,\,\,  .
\end{equation}
We used this formula and the estimated sky coverage to compute the
number counts using a ``high-purity'' sample with $p_s>5\, \sigma$,
containing 45 candidates, for which negligible contamination is
expected (Fig.~\ref{contam}).  After estimating the appropriate sky
coverage for this selection we computed the points reported in
Fig.\ref{lnls}.  We chose to bin the counts so as to have an equal
increase $\Delta N=5$ in the number of sources included in each bin,
when moving to fainter and fainter fluxes.  The error bars also include
the uncertainty in the measured $r_c$, on which the sky coverage
for each source depends.  Our result is compared to similar
measurements from the RDCS (Rosati et al. 1998) and 160SD (Vikhlinin
et al. 1998a) surveys, and to the predictions for an unevolving and
evolving XLF in the adopted $H_o=70$ km s$^{-1}$ Mpc$^{-1}$,
$\Omega_M=0.3$, $\Omega_{\Lambda} = 0.7$ cosmology.  The XLF evolution
is described using the same phenomenological $(A,B)$ model used in the
previous section, here using both the original parameters by Rosati et
al. (2000, dotted line) and the milder values by Mullis et al. (2004,
dashed line).  We also note that the no-evolution curve is slightly
higher than the one shown in Rosati et al. (2002), due to the
different XLF parameters used here (REFLEX vs. BCS).  The main result
from this plot is that the BMW-HRI and PSPC data agree very well with
each other and are globally consistent with a moderate
evolution of the XLF.  The observed counts can be taken first of all
as an {\it a posteriori} confirmation of the predicted low level of
contamination of the $p_s> 5\,\sigma$ sample, and of the self-consistency of the
computed sky coverage.  More substantially, the agreement with the
other surveys indicates that the bulk of the (faint) cluster
population is consistently detected by both PSPC- and HRI-based
surveys. However, one should keep in mind also the limitations of this
kind of plot.  In fact, the shape and amplitude of the logN-LogS is
mainly sensitive to the faint/intermediate range of the XLF (dominated
by low-redshift groups and clusters), while it is rather insensitive
to changes in the number of rare, high-luminosity clusters at high
redshift.

\section{Early follow-up results}
\label{deep}
\begin{figure*}
\includegraphics[width=16cm]{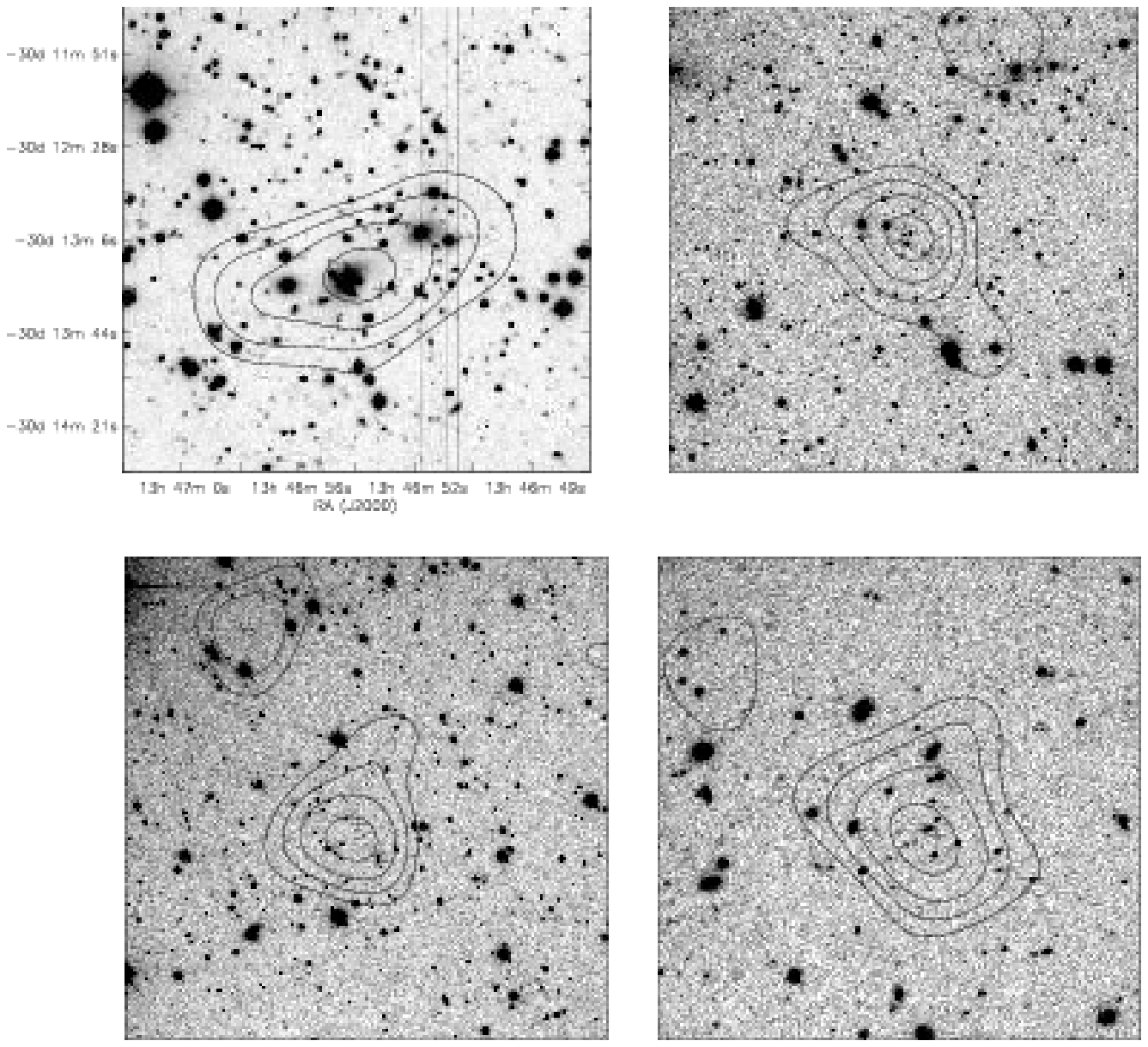}
\caption{Optical/X--ray overlays for a selection of new BMW--HRI groups/clusters at
different redshifts.  From top left to bottom right: BMW134654.9-301328,
$z_{spec}=0.358$; BMW212415.7-334754, $z_{spec}=0.92$; BMW122842.6-391612,
$z_{phot}\simeq 0.8$; BMW112059.0+130450, $z_{spec}=0.615$.   The
redshift for BMW212415.7-334754, a very compact cluster, awaits
further confirmation, being based on only 3 redshifts and displaying a
second system at $z=0.62$.
The X--ray images have been smoothed with a Gaussian filter ($\sigma = 40\arcsec$)
and the isophotes correspond to 0.7,1,2,3 standard deviations over the background
(after smoothing).   
Optical images are combined $Gunn$-$r+i$ $\sim 60\min$ total exposures with EFOSC2 at the
ESO 3.6~m telescope. For all but the first object the sides of the figures are 2.5\arcmin.  
}
\label{cl1226}
\end{figure*}

To provide a first hint of the general optical properties of BMW-HRI
clusters, we briefly summarize here some early results from the
ongoing follow-up campaign.  Our main cluster identification strategy
is currently based on multi-band imaging in the optical $g$, $r$, $i$
bands and near-infrared $J$, $H$, $K_s$ bands, with at least two,
typically three of these bands secured for each target.  A cluster is
considered as confirmed when a significant over-density of galaxies
with coherent colours is detected in the area of the X--ray source.
The current identification statistics, based on a total sample of 119
candidates observed so far, gives 83 confirmed, 19 rejected and 17
still uncertain clusters, in substantial agreement with the
contamination level estimated in this paper.  Photometric redshifts
for these objects are also estimated either approximately from the
mean colour of the cluster red sequence (when present), or more
accurately using the photo-$z$ codes of Fernandez-Soto et al. (2001),
and Bolzonella et al. (2000) when 3 or more bands are
available. A few of these clusters have also been observed
spectroscopically by us or have been found to be in common with other
surveys.  We cross-correlated our list with the latest version of the
160SD survey, with the unpublished RDCS catalogue and with the SHARC
(Romer et al. 2000) and WARPS (Perlman et al. 2002) published lists.
This comparison to PSPC-based surveys, performed early in the project,
provided us with encouraging confirmations, indicating that the survey
had the potential to efficiently peer into the high-redshift Universe.
Common high-redshift clusters include, for example: $\bullet$
BMW122657.3+333253 at $z=0.888$, first discovered by Cagnoni et
al. (2001) in the WGA survey and by Ebeling et al. (2001b) in the
WARPS survey;
$\bullet$ BMW052215.8-362452 at $z=0.53$, also 
found in the 160SD survey (Mullis et al. 2003); $\bullet$ two
(unpublished) RDCS clusters at $z=0.64$ and $z=0.808$ (Rosati, private
communication).  Our current list of new high-redshift clusters
includes another 7 confirmed objects reliably located at
$z_{phot}>0.6$ by $r$-$i$-$Ks$ photometric redshifts.  A few of these
are shown in Fig.~\ref{cl1226}.    These clusters will be the subject
of specific future papers.   X--ray/optical overlays and RGB images for
more BMW--HRI clusters can be seen at {\sl http://www.merate.mi.astro.it/$\sim$guzzo/BMW/gallery.html}.

\section{Summary and Conclusions}

We have presented in detail the construction of the BMW--HRI Cluster
Survey, a new sample of X--ray selected cluster candidates drawn from
the so-far poorly explored ROSAT HRI archive. 
We have shown that by selecting sources with detection significance larger than
$4\,\sigma$ and extension significance better than $\sim 3.5\,\sigma$,
one can obtain
a reliable sample of 154 cluster candidates with flux larger than $2
\times$10$^{-14}$ erg sec$^{-1}$ cm$^{-2}$ and  very interesting properties.  We
have performed extensive Monte Carlo simulations to recover the
survey selection function with respect to the major parameters
affecting the source detection.  These have allowed us to reconstruct
a sky coverage which is competitive with existing PSPC surveys over
the whole range of fluxes, with a particularly interesting solid angle
covered in the faintest bins.  These results show that, contrary to
expectations, the HRI data --- once they have been treated as discussed in C99,
i.e. eliminating the noisiest energy channels --- allow us to detect
low-surface-brightness objects out to high redshifts,
despite their higher instrumental background. We also estimated the expected sample
contamination due to background
fluctuations and false classifications, which turns out to be $27\%$
for this significance threshold.

We have used a high purity sample selected at $p_s>5\, \sigma$,
containing  45 objects expected to be virtually all true clusters, to
estimate the cluster number counts in the range $3\times$10$^{-14}$ to
$\sim 1 \times$10$^{-12}$ erg sec$^{-1}$ cm$^{-2}$.  This measurement is in
very good agreement with previous estimates from the RDCS and 160SD
surveys.

Possibly the most important aspect of the BMW--HRI cluster survey is that it will be 
the first large sample of clusters to be drawn from an instrument
independent from and with higher resolution than the ROSAT PSPC, on
which virtually all serendipitous cluster surveys so far have been
based\footnote{Notable recent exceptions are represented by new
samples being constructed using Chandra and XMM-Newton data, as done
respectively by Boschin (2002) and by the XMM-LSS survey (Valtchanov
et al. 2003).  In the first case, an accurate study of the survey
selection function, sky coverage and predicted redshift distribution
has been provided.  Unfortunately, no deep follow-up is being done
for this survey, where a significant fraction of clusters with
$z>1$ is expected. In the case of the XMM-LSS sample, on the other
hand, two $z\sim 1$ clusters have recently been discovered (Andreon,
priv. comm.).  However, no quantitative statistical description of the
survey sky coverage is available yet.}.  This means that the BMW--HRI
survey potentially includes some types of object which could have been missed in
previous surveys.  It is natural to expect that these are mostly small
groups, which the HRI is able to detect as extended.  Indeed,
we have specific examples in the survey of objects with core radii as
small as 6\arcsec.  The observed agreement among our number counts and
those from the RDCS and 160SD surveys, however, indicates that the
percentage of these small-sized objects does not seem to be sufficient
to represent a substantial incompleteness in PSPC surveys.  A finer
assessment of possible differences will be provided once the redshifts
for BMW--HRI clusters are measured, both by comparison of the faint end
slopes of the cluster XLF and of the number of clusters detected above
$z\sim 0.8$.  While the faint end of the number counts is dominated by
low-redshift, low-luminosity objects, the HRI resolution should begin to make
a difference when one goes to high redshifts.  The work of Ettori
et al. (2004), based on Chandra observations, seems to indicate that
$z>0.8$ clusters have on average a more compact profile than their
lower-redshift counterparts.  Keeping in mind that all our simulations
are based on the necessary but crude approximation of a $\beta=2/3$
profile, these findings would go in the direction of increasing the
probability of detecting clusters at high redshift, favored by the HRI high
spatial resolution.
This would probably be too mild to show a
significant effect in the integral number counts, yet might increase
the number of detectable $z\sim 1$ clusters by a significant factor.

\begin{acknowledgements}
We thank P. Rosati for for continuous encouragement and for allowing us to make a comparison with 
unpublished data from the RDCS, C. Mullis for cross-checks with the 160SD survey
and A. Vikhlinin for providing us with his number counts in
electronic form. We thank S. Borgani, H. B\"ohringer, I. Gioia, J.P. Henry and C. Mullis
for useful discussions and A. Finoguenov and W. Boschin for reading the manuscript.
We thank A. Misto for continuous data archiving assistance.
\end{acknowledgements}


\end{document}